\documentclass[twocolumn,aps,prb,showpacs,amsmath,amssymb]{revtex4}
\usepackage{graphicx,dcolumn,bm}
\usepackage{color}
\begin{document}

\title{Nonequilibrium steady states of the electric-field-driven Mott insulator:
Thermalization, emergent Wannier-Stark ladder, and dielectric breakdown}

\author{Woo-Ram Lee}
\author{Kwon Park}
\email{kpark@kias.re.kr}

\affiliation{School of Physics, Korea Institute for Advanced Study, Seoul 130-722, Korea}

\date{\today}

\begin{abstract}
In this work, we explore the possibility of emergent nonequilibrium steady states arising from the electric-field-driven Mott insulator via the Keldysh-Floquet dynamical mean field theory (DMFT), which can determine the fully-interacting, nonequilibrium steady-state Green's functions with the noninteracting counterparts as an input to the DMFT self-consistency loop. 
Unlike the retarded component, obtaining the lesser Green's function for the noninteracting system presents an important obstacle since the thermalization of the noninteracting system still requires a precise understanding of the dissipation mechanism.    
A crucial breakthrough in this work is that the noninteracting lesser Green's function can be determined in terms of the Wannier-Stark ladder (WSL) eigenstates, which are thermalized via the standard canonical ensemble according to the Markovian quantum master equation. 
As a result, it is shown that the intricate interplay between strong correlation and large electric field can generate a sequence of two dielectric breakdowns with the first induced by a coherent reconstruction of the mid-gap state within the Mott gap and the second by an incoherent tunneling through the biased Hubbard bands.
It is predicted that the reconstructed mid-gap state generates its own emergent WSL structure with a reduced effective electric field. 
The two dielectric breakdowns are mediated by a reentrant insulating phase, which is characterized by the population inversion, causing instability toward inhomogeneous current density states at weak electron-impurity scattering.
\end{abstract}


\maketitle
\section{Introduction}
\label{sec:introduction}

One of the most salient issues in modern physics is the quantum phase transition as a function of tuning parameters~\cite{Sachdev11}. 
Extensive studies have focused on the electron-electron correlation as a tuning parameter, which can induce quantum phase transitions to various emergent states of matter including the fractional quantum Hall state, high-temperature superconductor, and Mott insulator. 
With great advances in nanoscience~\cite{Mendez93, Sohn97} as well as cold atom physics~\cite{Raizen97, Bloch08, Giorgini08}, quantum phase transitions in nonequilibrium conditions, for example, via application of a large electric field~\cite{Kumai99, Taguchi00, Oka03, Oka05, Heidrich-Meisner10, Mierzejewski10, Han13, Asamitsu97, Turkowski05, Freericks06, Freericks08, Joura08, Tsuji08, Eckstein10, Amaricci12, Aron12a, Aron12b, Eckstein13, Mitra06, Okamoto08, Arrigoni13, Eckstein13a, Guenon13} or illumination of an intense radiation~\cite{Tsuji08, Zudov01, Mani02, Zudov03, Shi03, Durst03, Andreev03, Park04, Durst04, Oka09, Lindner11, Iwai03, Okamoto07, Lu12, Kiryukhin97, Miyano97, Fiebig98, Tsuji09, Tsuji11, Liu11, Liu12, Eckstein13b, Fausti11}, are attracting more and more intensive attentions. 
While a systematic understanding of the nonequilibrium quantum phase transition is elusive at present, studying the nonequilibrium steady state of the electric-field-driven Mott insulator, especially in two or higher dimensions~\cite{Freericks06, Freericks08, Joura08, Tsuji08, Eckstein10, Eckstein13, Amaricci12, Aron12a, Aron12b}, provides an important stepping stone to achieve such a goal, revealing a new possibility for emergent nonequilibrium steady states of matter, which are induced by strong correlation in a highly nonequilibrium situation.

Under the expectation that the Mott insulator eventually undergoes a phase transition(s) to metal at a sufficiently large electric field~\cite{Asamitsu97},
the central issue of the electric-field-driven Mott insulator is exactly how this phase transition, also known as the dielectric breakdown, should occur.  
One possibility is that the electric field induces some kind of charge transfer from one region of the system to another so that the Mott insulating region is effectively doped. 
We are interested in a more fundamental possibility that the dielectric breakdown occurs through the nonlinear response of the system keeping the spatial uniformity.
Constructing an accurate, microscopic theory for the nonlinear response, however, is a highly challenging problem since it requires going beyond the well-established linear response theory, which is rigorously formulated in terms of the Kubo formula.

To address this issue, in this work, we analyze the electric-field-driven Mott insulator by using the Keldysh-Floquet DMFT, which can capture both effects of strong correlation and large electric field with arbitrary strengths. 
The Floquet formulation of the Keldysh DMFT provides a convenient platform for the direct analysis of the steady state, which is contrasted to real-time formulations~\cite{Turkowski05, Freericks06, Freericks08, Eckstein10, Amaricci12, Eckstein13}, where the full time evolution of the system is considered along the Keldysh contour, requiring high computational costs to achieve the true steady-state limit. 
In our formulation, the fully-interacting Green's functions in the steady-state limit are directly connected with the noninteracting counterparts via the Keldysh-Dyson equation embedded in the Keldysh-Floquet DMFT self-consistency loop.

A crucial breakthrough in this work is that the noninteracting Green's functions are determined in terms of the exact energy eigenstates, called the WSL eigenstates. 
It is important to note that the total energy (which is the sum of the kinetic and the electric potential energy) is still conserved under a static electric field. 
The corresponding energy eigenstates of the noninteracting system are the WSL eigenstates, whose precise mathematical form is obtained in this work via exact diagonalization.

More importantly, it is shown by using the Markovian quantum master equation that the WSL eigenstates are individually thermalized via the standard canonical ensemble. 
The Markovian quantum master equation, or the {\it Lindblad} equation, is derived under the assumption that (i) the system is coupled to a boson bath, which is already in thermal equilibrium, (ii) the system-bath coupling is sufficiently weak so that the density matrix for the whole system can be approximated as a direct product between the individual density matrix of the system and bath, and (iii) the dynamics of the system forgets its past history rapidly, which is also known as the Markovian condition.
Determined by this thermalization scheme, which we call the WSL-wise thermalization,
the noninteracting lesser Green's function is plugged into the Keldysh-Floquet DMFT self-consistency loop as a key piece of the input information.

Now, it is worth mentioning that an essentially identical thermalization scheme to the above, where the WSL eigenstates are essential, has been used for the semiconductor superlattice problem~\cite{Calecki84, Lyanda-Geller95,Wacker02}, while not for the problem of the electric-field-driven Mott insulator. 
In early stages of the study on the latter problem, the role of the thermalization was not clearly understood so that many studies were devoted to the closed system, which is completely decoupled from the environment~\cite{Freericks06, Freericks08, Tsuji08, Eckstein10, Eckstein13}. 
A necessity of the proper thermalization for the open system was addressed initially by Joura and collaborators~\cite{Joura08}, who introduced an approximate ansatz on the density matrix.
Later, numerous efforts have been made to solve the Keldysh-Dyson equation in a direct manner for appropriately chosen system-bath coupling terms, which can be categorized into two classes: (i) coupling with an external fermion bath (which is not influenced by the electric field)~\cite{Han13, Tsuji09, Amaricci12, Aron12a, Aron12b}, and (ii) coupling with a boson bath composed of a single phonon mode~\cite{Eckstein13, Eckstein13a, Eckstein13b}.

The fermion-bath model is inspired by a mesoscopic transport problem of the small normal-metal ring penetrated by a time-dependent magnetic field (which induces an electromotive force along the circumference of the ring)~\cite{Buttiker85}. 
While providing reasonable dissipation effects, a fictitious fermion bath introduced in the fermion-bath model holds no direct relationship with the actual dissipating environment existing in condensed matter.
A more realistic bath model can be obtained via coupling with a boson bath, particularly, composed of phonons. 
Despite being more realistic, however, previous works using the boson-bath model suffer from a serious problem that the boson bath is composed of only a single phonon mode (for that matter, any finite number of modes), which, according to the Poincar\'{e} recurrence theorem~\cite{Bocchieri57, Percival61, Schulman78}, means that the system will return to the initial state eventually in a sufficiently long time, i.~e., there is no dissipation. 
To overcome this problem, one has to consider a boson bath composed of infinitely many phonon modes with a continuous energy spectrum.
In this work, we use the Caldeira-Leggett model, which represents the boson bath as a continuous set of infinitely many quantum harmonic oscillators. 
The Markovian quantum master equation is derived from the Caldeira-Leggett model with help of the Markovian condition combined with several other approximations.

As a result of the Keldysh-Floquet DMFT, which uses the noninteracting Green's functions as an input to the self-consistency loop, it is shown that the electric-field-driven Mott insulator undergoes a sequence of two phase transitions to metal with the first induced by a coherent reconstruction of the mid-gap state within the Mott gap and the second by an incoherent tunneling through the biased Hubbard bands. 
The reconstructed mid-gap state generates its own emergent WSL structure with a reduced effective electric field, as if strong correlation is renormalized away with exchange of the electric field.    
The two metallic phases are mediated by a reentrant insulating phase, which is characterized by the population inversion, causing instability toward inhomogeneous current density states at weak electron-impurity scattering.

The rest of the paper is organized as follows.
In Sec.~\ref{sec:Hamiltonian}, we present the precise mathematical form of the Hamiltonian in the dynamical vector potential gauge, which describes the Hubbard model under a static electric field.
The Hamiltonian also includes the electron-impurity interaction term.
In Sec.~\ref{sec:DMFT}, we explain the Keldysh-Floquet DMFT, which constitutes the main theoretical framework of this work. 
The Keldysh-Floquet DMFT self-consistency loop requires the knowledge of the noninteracting, retarded and lesser Green's function as an input to be complete. 
To determine the noninteracting lesser Green's function, we switch gears to the static scalar potential gauge. 
The noninteracting Hamiltonian is diagonalized exactly in Sec.~\ref{sec:WSL_eigenstate}, which provides the precise mathematical form of the WSL eigenstates. 
In Sec.~\ref{sec:thermalization}, we derive the Markovian quantum master equation to show that each WSL eigenstate is individually thermalized via the standard canonical ensemble for the energy dissipation mechanism described by the Caldeira-Leggett model.  
In Sec.~\ref{sec:conversion}, we determine the noninteracting lesser Green's function in the dynamical vector potential gauge by converting the WSL-wise thermalization from the static scalar to the dynamical vector potential gauge. 
In Sec.~\ref{sec:KeldyshDyson}, we provide the concrete mathematical form of the Keldysh-Floquet DMFT self-consistency loop.
In Sec.~\ref{sec:IPT}, we discuss the validity of the iterated perturbation theory (IPT), which is the impurity solver of this work. 
It is shown that the IPT becomes exact in both limits of weak and strong $U$ even within the Keldysh-Floquet DMFT.

Results are discussed in Sec.~\ref{sec:results}.
In Sec.~\ref{sec:local_spectrum}, we discuss the evolution of the nonequilibrium steady states in terms of the local density of states (DOS), occupation number and distribution function as a function of electric field. 
In Sec.~\ref{sec:DC_current}, we compute the direct-current density by using an exact formula, which relates the direct-current density with the (fully-interacting) lesser Green's function. 
To provide further physical insights, we compare the results obtained from the exact formula with those from the tunneling formula.
It is shown that the tunneling formula can be derived rigorously in the weak-tunneling limit.
Finally, we conclude in Sec.~\ref{sec:conclusion}.

\section{Hamiltonian}
\label{sec:Hamiltonian}

We begin our analysis by writing the Hamiltonian for the Hubbard model under a static electric field in the dynamical vector potential gauge, $\phi = 0$ and $\mathbf{A} = -c\mathbf{E}t$:
\begin{align}
H = &-\sum_{\langle ij \rangle, \sigma} t_{ij} \left[ e^{i\varphi_{ij}(t)} c_{i\sigma}^\dag c_{j\sigma} +\textrm{H. c.}\right] \nonumber\\
&+U \sum_i n_{i\uparrow}  n_{i\downarrow}  +V \sum_{i,\sigma} n_{i\sigma} n_{i,\mathrm{imp}}, 
\label{eq:Hamiltonian}
\end{align}
where  $t_{ij}$ is the hopping amplitude and $c_{i\sigma}^\dag$ $(\sigma =\uparrow, \downarrow)$ is the electron creation operator for the $i$-th lattice site.
$n_{i\sigma}$ and $n_{i, {\rm imp}}$ are the electron and impurity number operator, respectively.  
$U$ ($>0$) and $V$ are the on-site electron-electron and electron-impurity interaction strength, respectively.
$\varphi_{ij}$ denotes the Peierls phase accumulated along a path connecting between the $i$-th and $j$-th site: 
$\varphi_{ij}(t) = \frac{|e|}{\hbar c} \int_{\mathbf{r}_i}^{\mathbf{r}_j} \mathbf{A}(t) \cdot d\mathbf{r}$.

In addition to the $U$ and $V$ term in Eq.~\eqref{eq:Hamiltonian}, there is an additional system-bath coupling term, which is responsible for the thermalization of the system.
We provide the concrete mathematical form for this system-bath coupling term in Sec.~\ref{sec:thermalization}, which gives rise to the thermalization of the noninteracting system via the Markovian quantum master equation. 
With the information on the thermalization of the noninteracting system plugged in as an input, the thermalization of the fully-interacting system can be determined self-consistently via the Keldysh-Dyson equation, which is embedded in the Keldysh-Floquet DMFT self-consistency loop.

In this work, we focus on nonequilibrium steady states with the spatially-uniform charge density. 
It is shown in Sec.~\ref{sec:DC_current} that such nonequilibrium steady states are inevitably accompanied by the corresponding steady-state direct current, ${\bf J}$, if electrons are allowed to scatter with themselves and/or impurities via the $U$ and $V$ term in the Hamiltonian. 
(It is shown that no direct currents exist in the absence of such elastic scatterings. 
See Sec.~\ref{sec:exact_formula} for details.) 
With the direct current maintained via the steady flow of electrons from/to an external reservoir, or battery, any heat generated by the steady-state direct current, i.~e., ${\bf E} \cdot {\bf J}$, is transferred to the bath through the dissipation mechanism governed by the Markovian quantum master equation, which is derived in Sec.~\ref{sec:thermalization}.
The temperature of the system (which is dictated by the bath via the Markovian quantum master equation) is maintained under the assumption that the bath is thermalized by itself and completely uninfluenced by the system.  
Finally, throughout this work, the charge density is set to be at half filling.

\section{Keldysh-Floquet dynamical mean field theory}
\label{sec:DMFT}

Being translationally invariant in the dynamical vector potential gauge, the Hamiltonian in Eq.~\eqref{eq:Hamiltonian} can be addressed by the DMFT, which is known to provide an accurate description of the Mott insulator~\cite{Georges96,Imada98}.
There is, however, a price to be paid so that the Hamiltonian now becomes explicitly time-dependent.  
Fortunately, the DMFT can be reformulated via the Keldysh formalism to address the time-dependent Hamiltonians~\cite{Freericks06, Freericks08}.
A further reformulation of the Keldysh DMFT in terms of the Floquet representation is proven useful for studying the steady states~\cite{Joura08, Tsuji08, Tsuji09}.  
For convenience, the electric-field-driven Hubbard model is solved in the hypercubic lattice (where the DMFT becomes exact in the limit of large spatial dimensions, $d \rightarrow\infty$) with the electric field applied along the diagonal direction, in which case $t_{ij}$ and $|\mathbf{E}|$ are set to scale as $t^*/2\sqrt{d}$ and $E^* \sqrt{d}$, respectively~\cite{Freericks06}.
We call $t^*$ and $E^*$ the normalized hopping amplitude and electric field strength, respectively.

\begin{figure}
\centering
\includegraphics[width=0.45\textwidth]
{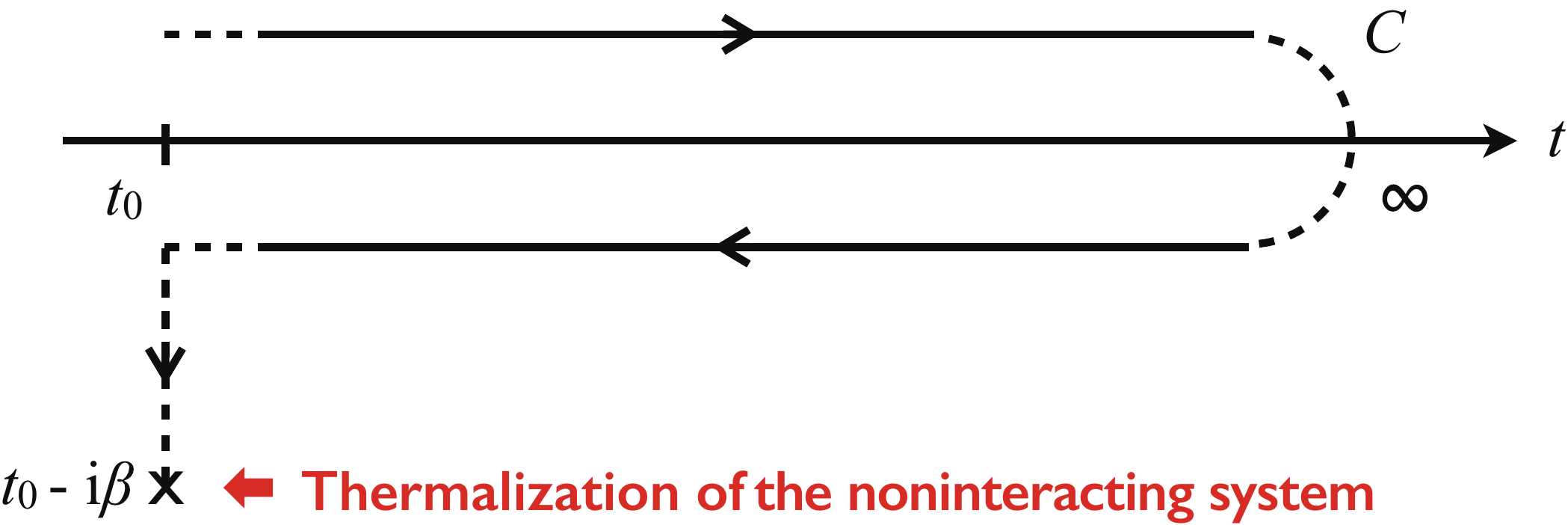} \\
\caption{
Schematic diagram for the Keldysh contour. 
The fully-interacting Green's functions are connected with the noninteracting counterparts at the initial time, $t_0$, at which the interaction between electrons is switched on.   
As conventional, $t_0$ is assumed to be in the infinite past so that we can focus on the steady-state behaviors of the system.   
The thermalization of the noninteracting system is determined schematically at $t_0-i\beta$, where $\beta=1/k_{\rm B}T$ with $T$ being the temperature.
The actual determination of the thermalization for the noninteracting system is accomplished via the Markovian quantum master equation, which is derived in Sec.~\ref{sec:thermalization}.
} 
\label{Fig1_KeldyshContour}
\end{figure}

Our Keldysh-Floquet DMFT is implemented, based on the standard Keldysh-contour expansion~\cite{Rammer86, Haug08}, along which the fully-interacting Green's functions are connected with the noninteracting counterparts in the infinite past, at which the interaction between electrons is switched on.   
Figure~\ref{Fig1_KeldyshContour} shows a schematic diagram for the Keldysh contour.
It is important to note that, here, the electric field is assumed to be always on, or turned on even prior to the switching of the interaction between electrons.
In this scheme, the thermalization of the fully-interacting system can be determined self-consistently via solving the Keldysh-Dyson equation with that of the noninteracting system plugged in as an input to the DMFT self-consistency loop.    
The thermalization of the noninteracting system is, in turn, determined via solving the Markovian quantum master equation, whose derivation is postponed to Sec.~\ref{sec:thermalization}.

\begin{figure*}
\centering
\includegraphics[width=0.9\textwidth]
{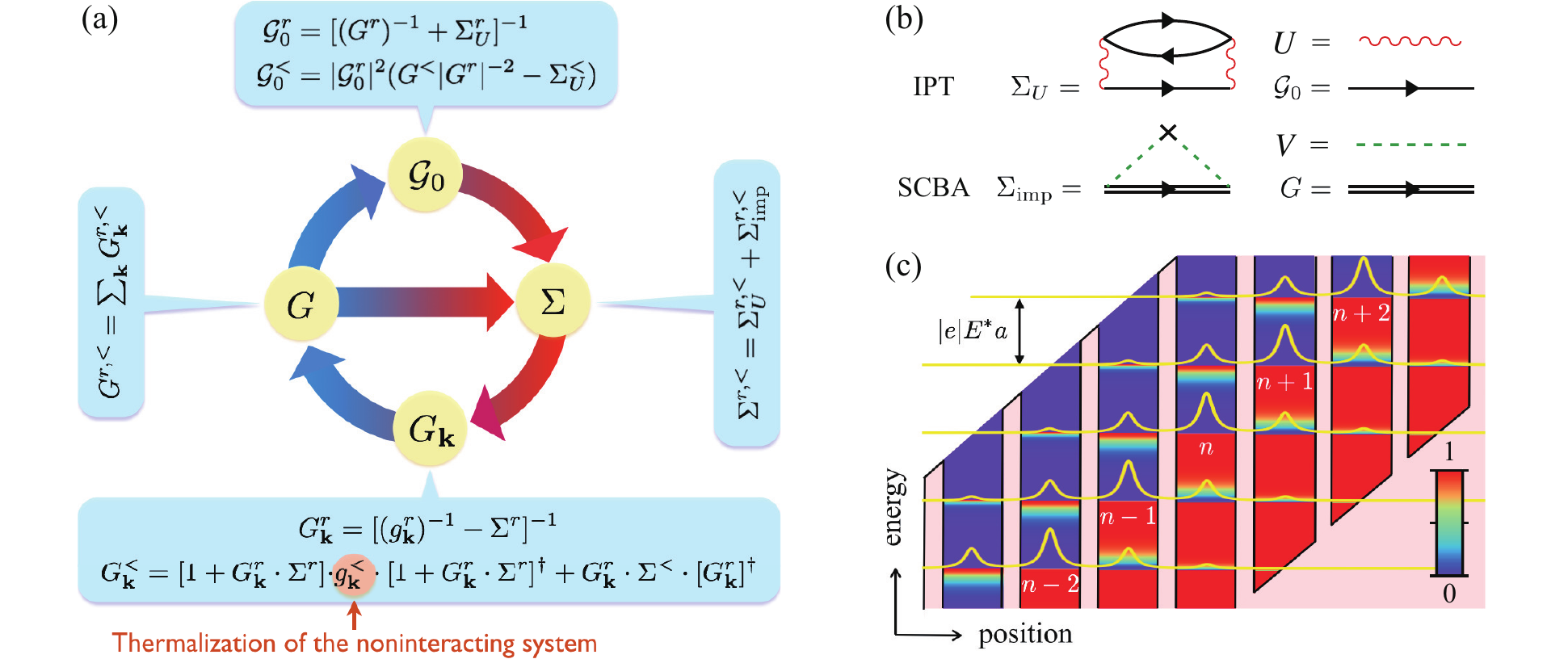} \\
\caption{
(a) Self-consistency loop for the Keldysh-Floquet DMFT. 
Each object in the yellow circle is determined by the equations in the corresponding blue rounded box. 
$\mathcal{G}_0$ and $G$ denote the Weiss and local Green's function, respectively.
The full lattice Green's function, $G_\mathbf{k}$, is connected with the noninteracting counterpart, $g_\mathbf{k}$, via the Keldysh-Dyson equation. 
The self-energy, $\Sigma$, has two contributions with one originating from electron-electron scattering, $\Sigma_U$, and the other from electron-impurity scattering, $\Sigma_\mathrm{imp}$. 
Here, all Green's functions as well as self-energies are represented as Floquet matrices.
See Sec.~\ref{sec:KeldyshDyson} for details.
(b) Feynman diagram for the self-energy. 
$\Sigma_U$ is computed via the DMFT impurity solver called the iterated perturbation theory (IPT)~\cite{Georges96}.
Similar to equilibrium, the IPT is regarded as an interpolation scheme connecting between both limits of weak and strong $U$, where $\Sigma_U$ becomes exact even in the presence of a static electric field.
$\Sigma_\mathrm{imp}$ is computed via the self-consistent Born approximation (SCBA).
See Sec.~\ref{sec:KeldyshDyson} and \ref{sec:IPT} for details.
(c) Wannier-Stark ladderwise thermalization. 
Completing the self-consistency loop, $g^<_\mathbf{k}$ contains the crucial information on the thermalization of noninteracting system in terms of the WSL eigenstates, whose wave function profiles are plotted schematically in yellow lines with each site represented as a quantum well.
The local distribution function for the noninteracting system is shown in color scale (with red being full and blue being empty) as a function of energy, which is different from the usual Fermi-Dirac distribution function due to the fact that the WSL eigenstates are spread over multiple sites, while thermalized as a whole. 
See Sec.~\ref{sec:thermalization} for details.
} 
\label{Fig2_SC_loop}
\end{figure*}

Based on the above Keldysh-contour expansion, the self-consistency loop for the Keldysh-Floquet DMFT can be constructed concretely. 
Figure~\ref{Fig2_SC_loop}~(a) provides a detailed flow chart for the self-consistency loop, which can determine the fully-interacting retarded and lesser Green's function, $G^{r,<}_\mathbf{k}$, from the noninteracting counterparts, $g^{r,<}_\mathbf{k}$, via the Keldysh-Dyson equation~\cite{Comment_G}.
Similar to equilibrium, we begin with an initial guess for the Weiss functions, $\mathcal{G}^{r,<}_0$, and continue until all functions are converged within numerical tolerance~\cite{Comment_LM}. 
After convergence, the local density of states (DOS) and occupation number can be obtained, respectively, as follows: 
\begin{align}
\rho_\mathrm{loc}(\omega+n\Omega) &= -\frac{1}{\pi}\mathrm{Im} G^r_{nn}(\omega), 
\\ 
N_\mathrm{loc}(\omega+n\Omega) &= \frac{1}{2\pi}\mathrm{Im} G^<_{nn}(\omega), 
\end{align}
where $\hbar \Omega=|e| E^* a$  with $a$ being the lattice constant.
The local distribution function is defined as
\begin{align}
f_\mathrm{loc}(\omega)=N_\mathrm{loc}(\omega)/\rho_\mathrm{loc}(\omega),
\end{align} 
which reduces to the usual Fermi-Dirac distribution function in the low-field limit.

Obtaining $g^{r}_\mathbf{k}$ is straightforward in the sense that it only requires solving noninteracting equations of motion.
A challenging part is to determine $g^{<}_\mathbf{k}$, which is crucial for the thermalization of the system. 
To determine $g^{<}_\mathbf{k}$, it is convenient, for the time being, to switch gears to the static scalar potential gauge, where it is clear that the total energy, i.~e., the sum of the kinetic and electric potential energy, is conserved under a static electric field (at least, during the time scale set by the inverse width of the energy level broadened via inelastic scattering).
In the following section, we perform exact diagonalization of the noninteracting Hamiltonian in the static scalar potential gauge to show that the corresponding energy eigenstates are given by the WSL eigenstates with the eigenvalue $n\hbar\Omega=n|e|E^*a$, where $n$ is the WSL index denoting the central site with the maximum weight.
The wave function profiles for the WSL eigenstates are shown in Fig.~\ref{Fig2_SC_loop}~(c) schematically.

\subsection{Diagonalization of the noninteracting system: Wannier-Stark ladder eigenstates}
\label{sec:WSL_eigenstate}

In this section, we perform exact diagonalization of the noninteracting Hamiltonian in the static scalar potential gauge. 
The noninteracting Hamiltonian is written in the static scalar potential gauge as follows:
\begin{align}
H_\mathrm{stat} =& H^K_\mathrm{stat} +H^V_\mathrm{stat} ,
\end{align}
where
\begin{align}
H^K_\mathrm{stat} =& -\sum_{\langle ij \rangle} t_{ij} \left( c_i^\dag c_j +\textrm{H. c.} \right) 
=\sum_\mathbf{k} \epsilon_\mathbf{k} c^\dag_\mathbf{k}  c_\mathbf{k}  ,
\\
H^V_\mathrm{stat} =& |e| \mathbf{E} \cdot \sum_{i} \mathbf{r}_i c^\dagger_i c_i ,
\end{align}
where the spin index is dropped for simplicity. 
In the above, $\epsilon_\mathbf{k}= -\frac{t^*}{\sqrt{d}}\sum_{j=1}^d \cos{k_j a}$, and $\mathbf{r}_i$  is the position vector for the $i$-th site.
As mentioned before, here, we assume that the electric field is applied along the diagonal direction of the hypercubic lattice:
$\mathbf{E} = E^* \sum_{j=1}^d \hat{\bf e}_j$.

Since the momentum is conserved along all directions perpendicular to the electric field, 
it is convenient  to decompose $\mathbf{k}$ into the perpendicular, $\mathbf{k}_\perp$, and the diagonal, $\mathbf{k}_{\rm d}$, component. 
By using this decomposition, $H^K_{\rm stat}$ can be rewritten as follows: 
\begin{align}
&H^K_\mathrm{stat} 
= \sum_{\mathbf{k}_\perp, \mathbf{k}_{\rm d}} \epsilon_{{\mathbf{k}_\perp+\mathbf{k}_d}} c^\dag_{\mathbf{k}_\perp, \mathbf{k}_{\rm d}}  c_{\mathbf{k}_\perp,\mathbf{k}_{\rm d}}  
\nonumber \\
&=\sum_{\mathbf{k}_\perp, k_{\rm d}} 
\left[
\epsilon_{\mathbf{k}_\perp} \cos{(k_{\rm d}/\sqrt{d})}
-\bar{\epsilon}_{\mathbf{k}_\perp} \sin{(k_{\rm d}/\sqrt{d})}
\right]
 c^\dag_{\mathbf{k}_\perp, k_{\rm d}}  c_{\mathbf{k}_\perp, k_{\rm d}}  
 \nonumber \\
&=\sum_{\mathbf{k}_\perp, k_{\rm d}} 
\zeta_{\mathbf{k}_\perp} \cos{(k_{\rm d}/\sqrt{d}  +\theta_{\mathbf{k}_\perp})}
 c^\dag_{\mathbf{k}_\perp, k_{\rm d}}  c_{\mathbf{k}_\perp, k_{\rm d}}  , 
 \label{K_stat_kd}
 \end{align}
where $\bar{\epsilon}_\mathbf{k} = - \frac{t^*}{\sqrt{d} }\sum_{j=1}^d \sin{k_j a}$,
$\zeta_\mathbf{k}=\sqrt{\epsilon^2_\mathbf{k} +\bar{\epsilon}^2_\mathbf{k}}$, 
$\tan{\theta_{\bf k}}=\bar{\epsilon}_{\bf k}/\epsilon_{\bf k}$,
and $k_{\rm d}=|\mathbf{k}_{\rm d}|$.
Similarly, $H^V_{\rm stat}$ can be rewritten as follows:
\begin{align}
H^V_\mathrm{stat} 
= \hbar \Omega \sum_{\mathbf{k}_\perp, i_\mathrm{d}} i_\mathrm{d} c^\dagger_{\mathbf{k}_\perp, i_{\rm d}} c_{\mathbf{k}_\perp, i_{\rm d}} ,
\label{V_stat}
\end{align}
where $i_\mathrm{d} = \frac{1}{a} \sum_{j=1}^d \mathbf{r}_i \cdot \hat{\bf e}_j$ is the diagonal-plane index, which is related to $k_{\rm d}$ via Fourier transformation. 
Now, in order to represent both $H^K_{\rm stat}$ and $H^V_{\rm stat}$ via the same parameter set, we perform a partial Fourier transformation of Eq.~\eqref{K_stat_kd} from the space of $k_{\rm d}$ to $i_{\rm d}$:
\begin{align}
H^K_\mathrm{stat} 
=& \sum_{\mathbf{k}_\perp} \frac{\zeta_{\mathbf{k}_\perp}}{2}  \sum_{ i_{\rm d} }  
\left[ 
e^{i\theta_{\mathbf{k}_\perp}} c^\dag_{\mathbf{k}_\perp, i_{\rm d}+1}  c_{\mathbf{k}_\perp, i_{\rm d}}  +\textrm{H. c.} 
\right] .
\label{K_stat_id}
\end{align}
Combining Eqs.~\eqref{V_stat} and \eqref{K_stat_id} gives rise to the Hamiltonian in the following form:  
\begin{align}
H_\mathrm{stat} 
= & \sum_{\mathbf{k}_\perp} \frac{\zeta_{\mathbf{k}_\perp}}{2}  \sum_{ i_{\rm d} }  
\left[ 
e^{i\theta_{\mathbf{k}_\perp}} c^\dag_{\mathbf{k}_\perp, i_{\rm d}+1}  c_{\mathbf{k}_\perp, i_{\rm d}}  +\textrm{H. c.} 
\right] 
\nonumber \\
&+\hbar \Omega \sum_{\mathbf{k}_\perp, i_\mathrm{d}} i_\mathrm{d} c^\dagger_{\mathbf{k}_\perp, i_{\rm d}} c_{\mathbf{k}_\perp, i_{\rm d}} ,
\label{H_stat}
\end{align}
which is analogous to the Hamiltonian for a semiconductor superlattice~\cite{Wacker02}. 
Motivated by this analogy, we consider the following unitary transformation:
\begin{align}
c_{\mathbf{k}_\perp, i_{\rm d}}^\dag = \sum_n J_{n-i_{\rm d}}(\zeta_{\mathbf{k}_\perp} / \hbar\Omega) 
e^{- i \theta_{\mathbf{k}_\perp} i_{\rm d}} \tilde{c}_{\mathbf{k}_\perp, n}^\dag ,
\label{unitary}
\end{align}  
where $J_n(x)$ is the Bessel function of the first kind.
Plugging Eq.~\eqref{unitary} to \eqref{H_stat} diagonalizes the Hamiltonian exactly:
\begin{align}
H_\mathrm{stat} = \hbar\Omega \sum_{\mathbf{k}_\perp, n} n ~  \tilde{c}_{{\mathbf{k}_\perp}, n}^\dag \tilde{c}_{{\mathbf{k}_\perp}, n},
\label{HamiltonianWS}
\end{align}
where we have used the following sum rules for the product of Bessel functions~\cite{Bevilacqua11}:
\begin{align} 
\sum_j J_{n-j}(x) J_{n'-j}(x) 
&= \delta_{nn'} ,
\\
\sum_j j J_{n-j}(x) J_{n'-j}(x) 
&= n \delta_{nn'} - \frac{x}{2} (\delta_{n,n'+1} + \delta_{n,n'-1}) .
\end{align} 
With the energy eigenvalue precisely given by $n\hbar\Omega$, the explicit wave function for the corresponding eigenstates, called the WSL eigenstates~\cite{Wannier60}, is written as follows:
\begin{align}
\psi^{\rm WSL}_{\mathbf{k}_\perp, n} (i_{\rm d}) \equiv 
\langle 0| c_{\mathbf{k}_\perp, i_{\rm d}} 
\tilde{c}^\dagger_{\mathbf{k}_\perp, n} |0\rangle 
= J_{n-i_{\rm d}}(\zeta_{\mathbf{k}_\perp} / \hbar\Omega) 
e^{i \theta_{\mathbf{k}_\perp} i_{\rm d}} ,
\label{WSLstate}
\end{align}
which indicates that the $n$-th WSL state is localized around the $n$-th site.

Now, it is important to check if the above results are consistent with those obtained in the dynamical vector potential gauge. 
To this end, let us compute the local DOS, which should be gauge-invariant.
In the static scalar potential gauge, we begin by computing the local probability weight for an electron belonging to a given WSL eigenstate, say, the $l$-th, to exist at a fixed diagonal plane, say, $i_{\rm d}=0$:
\begin{align} 
P_l(i_{\rm d}=0)
&=\sum_{\mathbf{k}_\perp} |\psi^{\rm WSL}_{\mathbf{k}_\perp, l} (i_{\rm d}=0)|^2 
\nonumber \\
&= \sum_{\mathbf{k}_\perp} J^2_{l}(\zeta_{\mathbf{k}_\perp} / \hbar\Omega)
\nonumber \\
&=  \sum_{\mathbf{k}_\perp, \mathbf{k}_{\rm d}} J^2_{l}(\zeta_{\mathbf{k}_\perp+\mathbf{k}_{\rm d}} / \hbar\Omega) ,
\label{P_l_k_sum}
\end{align}
where the last line is obtained since 
$\zeta_{\mathbf{k}_\perp +\mathbf{k}_{\rm d}}=\zeta_{\mathbf{k}_\perp}$ for arbitrary $\mathbf{k}_{\rm d}$. 
With the summation over $\mathbf{k}$ transformed to the integral via $\sum_\mathbf{k}  = \int d^d (ka)/(2\pi)^d$,
Eq.~\eqref{P_l_k_sum} is rewritten as follows:
\begin{align}
P_l(i_{\rm d}=0)
&= \sum_{\mathbf{k}} J^2_{l}(\zeta_{\mathbf{k}} / \hbar\Omega)
\nonumber \\
&= \int \frac{d^d (ka)}{(2\pi)^d} J^2_{l}(\zeta_{\mathbf{k}} / \hbar\Omega)
\nonumber \\
&= 2\pi\int^{\infty}_{0} d\zeta \; \zeta \rho_{\rm non}(\zeta) J^2_{l}(\zeta/\hbar\Omega)
\nonumber \\
&= e^{-\frac{1}{2}(t^*/\hbar\Omega)^2} I_l ((t^*/\hbar\Omega)^2/2 ) ,
\label{P_l}
\end{align} 
where $I_l(x)$ is the modified Bessel function of the first kind. 
In the above, $\rho_{\rm non}$ is called the noninteracting joint DOS, which is computed to be $\rho_{\rm non}(\zeta)=e^{-(\zeta/t^*)^2}/(\pi t^{*2})$ for the hypercubic lattice in the limit of infinite spatial dimensions~\cite{Turkowski05,Tsuji08}. 
See Appendix~\ref{appen:jointDOS} for details of the computation of the noninteracting joint DOS.

Based on the fact that $l\hbar\Omega$ is the energy eigenvalue of the $l$-th WSL eigenstate, the local DOS is then determined as follows:
\begin{align} 
\rho_{\rm loc}(\omega)
&=\sum_l P_l(i_{\rm d}=0) \delta(\hbar\omega-l\hbar\Omega) 
\nonumber \\
&=\sum_l e^{-\frac{1}{2}(t^*/\hbar\Omega)^2} I_l ((t^*/\hbar\Omega)^2/2 ) \delta(\hbar\omega-l\hbar\Omega) ,
\end{align} 
which is precisely identical to the local DOS obtained in the dynamical vector potential gauge previously~\cite{Tsuji08}. 
For the sake of convenience, we recapitulate the computation of the local DOS in the dynamical vector potential gauge below. 
Before doing so, we extend the local DOS as follows with a finite level broadening:
\begin{align} 
\rho_{\rm loc}(\omega)
=\sum_l e^{-\frac{1}{2}(t^*/\hbar\Omega)^2} I_l ((t^*/\hbar\Omega)^2/2 ) \rho_{{\rm lad}, l}(\omega) ,
\label{LDOS_stat}
\end{align} 
where
\begin{align} 
\rho_{{\rm lad}, l}(\omega)
\equiv -\frac{1}{\pi} \mathrm{Im} g^r_{{\rm lad}, l} (\omega) = -\frac{1}{\pi} \mathrm{Im} \frac{1}{\hbar\omega-l\hbar\Omega+i\Gamma/2},
\end{align} 
where $\Gamma$ denotes the level broadening width.

In the dynamical vector potential gauge, the local DOS can be obtained from the imaginary part of the retarded local Green's function:
\begin{align}
\rho_{\rm loc}(\omega+n\Omega) = -\frac{1}{\pi} \mathrm{Im} \sum_{\bf k} \left[ g^r_{\bf k}(\omega) \right]_{nn} ,
\end{align}
where the noninteracting retarded Green's function, $g^r_{\bf k}(\omega)$, is written in the Floquet representation as follows~\cite{Tsuji08}:
\begin{align}
g_\mathbf{k}^r(\omega) 
&= \mathcal{U}_\mathbf{k} \cdot  g_\mathrm{lad}^r(\omega) \cdot  \mathcal{U}_\mathbf{k}^\dag , 
\label{RetWSL}
\end{align}
where $(\mathcal{U}_\mathbf{k})_{mn}=J_{m-n}(\zeta_\mathbf{k}/\hbar\Omega) e^{i(m-n)\theta_\mathbf{k}}$ and
$(g_\mathrm{lad}^r)_{mn}(\omega)$ $= (\hbar\omega +n\hbar\Omega  +i\Gamma/2)^{-1}\delta_{mn}$
with $-\Omega/2 < \omega \leq \Omega/2$. 
Here, it is important to note that, setting aside an unimportant overall phase factor (and the fact that ${\bf k}$ covers the whole range, not confined to ${\bf k}_\perp$, which becomes unimportant after the ${\bf k}$-summation), $({\cal U}_{\bf k})_{mn}$ is precisely identical to the wave-function amplitude of the $n$-th WSL state at the $m$-th site in Eq.~\eqref{WSLstate}.
Eventually, this fact leads to the conclusion that the summation over ${\bf k}$ produces exactly the same local DOS as that in the static scalar potential gauge.
Specifically,
\begin{align}
\rho_{\rm loc}(\omega+n\Omega) &= -\frac{1}{\pi} \mathrm{Im} \sum_{\bf k} \left[ g^r_{\bf k}(\omega) \right]_{nn} 
\nonumber \\
&=-\frac{1}{\pi} \mathrm{Im} \sum_{\bf k} \sum_q \frac{[J_{n-q}(\zeta_{\bf k}/\hbar\Omega)]^2}{\hbar\omega+q\hbar\Omega+i\Gamma/2}
\nonumber \\
&=-\frac{1}{\pi} \mathrm{Im} \sum_{\bf k} \sum_l \frac{[J_l(\zeta_{\bf k}/\hbar\Omega)]^2}{\hbar(\omega+n\Omega)-l\hbar\Omega+i\Gamma/2} ,
\end{align}
which reduces to the following after the transformation from the Floquet (where the frequency appears in the form of $\omega+n\Omega$ with $-\Omega/2 < \omega \leq \Omega/2$) to the Wigner representation (where $\omega$ is not restricted):
\begin{align}
\rho_{\rm loc}(\omega) = \sum_l e^{-\frac{1}{2}(t^*/\hbar\Omega)^2} I_l ((t^*/\hbar\Omega)^2/2 ) \rho_{{\rm lad}, l}(\omega),
\end{align}
which is precisely identical to Eq.~\eqref{LDOS_stat}.

Since the WSL eigenstates are true energy eigenstates, it is reasonable to expect that electrons are thermalized in terms of the WSL eigenstates, satisfying the standard canonical ensemble. 
In the following section, we prove this to be indeed the case by deriving the Markovian quantum master equation via the standard procedure of integrating out bath degrees of freedom from the Caldeira-Leggett model~\cite{Breuer02}, which is then solved in the steady-state limit.

\subsection{Thermalization of the Wannier-Stark ladder eigenstates: Markovian quantum mater equation}
\label{sec:thermalization}

\subsubsection{Caldeira-Leggett model}
\label{sec:Caldeira-Leggett}

To describe the dynamics of an open system, whose energy is dissipated through the system-bath coupling, we follow a seminal work by Caldeira and Leggett~\cite{Leggett87}, where the system is coupled to a bosonic thermal bath.
Specifically, we set up the total Hamiltonian, $H$, which is composed of the system Hamiltonian, $H_{\rm S}$, the bath Hamiltonian, $H_{\rm B}$, and the interaction Hamiltonian for the system-bath coupling, $H_{\rm I}$: $H = H_{\rm S} + H_{\rm B} + H_{\rm I}$.

First, the system Hamiltonian is written as follows:
\begin{align}
H_{\rm S} = \frac{\zeta}{2} \sum_{j=-\infty}^\infty \left( e^{i\theta} c^\dagger_{j+1} c_j + \textrm{H. c.} \right)  +\frac{\hbar\Omega}{a} X, 
\label{H_S}
\end{align}
which is exactly the same Hamiltonian as in Eq.~\eqref{H_stat} in Sec.~\ref{sec:WSL_eigenstate} except that the perpendicular momentum index, ${\bf k}_\perp$, is dropped here for the sake of convenience. 
It is shown in Sec.~\ref{sec:master_equation} that contributions from various ${\bf k}_\perp$ can be summed independently at the end of the calculation. 
The center-of-mass position operator, $X$, is defined by $X = a \sum_j j c^\dagger_j c_j$.

Second, following Caldeira and Leggett, we assume that the bath is described by a set of harmonic oscillators, in which case the bath Hamiltonian is given as follows:
\begin{align}
H_{\rm B} = \sum_{\alpha = 1}^N \bigg(\frac{p_\alpha^2}{2m_\alpha} + \frac{1}{2} m_\alpha \omega_\alpha^2 x_\alpha^2 \bigg) 
= \sum_{\alpha = 1}^N \hbar\omega_\alpha b_\alpha^\dag b_\alpha,
\end{align}
where $x_\alpha$ and $p_\alpha$ are the position and momentum operator of the $\alpha$-th harmonic oscillator mode, respectively. 
As usual, $b_\alpha^\dag$ and $b_\alpha$ are the creation and annihilation operator.

Finally, the system-bath coupling is described by the following interaction Hamiltonian:
\begin{align}
H_{\rm I} = - \sum_{\alpha=1}^N C_\alpha x_\alpha X,
\end{align}
where $C_\alpha$ is the coupling strength to the $\alpha$-th harmonic oscillator mode. 
We are interested in the weak-coupling limit, $C_\alpha \ll 1$, where the density matrix, $\hat{\rho}$, for the total system can be approximated as the direct product between that of the system, $\hat{\rho}_{\rm S}$, and bath, $\hat{\rho}_{\rm B}$:
$\hat{\rho} \simeq \hat{\rho}_{\rm S} \otimes \hat{\rho}_{\rm B}$,
which is known as the Born approximation.
Combined with other approximations explained in detail in the following section, 
the interaction Hamiltonian eventually leads to the Markovian quantum master equation.

As shown in Sec.~\ref{sec:WSL_eigenstate}, the system Hamiltonian can be exactly diagonalized by using the following unitary transformation between the creation/annihilation operator in the position basis to those in the WSL eigenbasis:
\begin{align}
c_j^\dag = \sum_n J_{n-j}(\zeta / \hbar\Omega) e^{- i \theta j} \tilde{c}_n^\dag ,
\label{unitary2}
\end{align} 
by which the system Hamiltonian is rewritten as follows:
\begin{align}
H_{\rm S} = \hbar\Omega \sum_n n \tilde{c}^\dagger_n \tilde{c}_n .
\end{align} 
Similarly, the interaction Hamiltonian is rewritten in the WSL eigenbasis as follows: 
\begin{align}
H_{\rm I} 
= & \sum_{\alpha = 1}^N C_\alpha \sqrt{\frac{\hbar}{2m_\alpha\omega_\alpha}} (b_\alpha + b_\alpha^\dag) \nonumber\\
& \times a \sum_{n} 
\left[ 
\frac{\zeta}{\hbar\Omega} ( \tilde{c}^\dagger_{n+1} \tilde{c}_n + \tilde{c}^\dagger_{n} \tilde{c}_{n+1} ) 
-n\tilde{c}^\dagger_n \tilde{c}_n
\right]. 
\label{Coupling}
\end{align}
It is shown in the following section that the first term of the summand with respect to $n$ is the main term responsible for the relaxation of the WSL eigenstates via the bosonic thermal bath.
The second term causes the decoherence of the WSL eigenstates.

\subsubsection{Derivation of the Markovian quantum master equation}
\label{sec:master_equation}

The time evolution of the density operator, $\hat{\rho}(t)$, for the total system is governed by the Liouville equation, which is given in the interaction picture as follows: 
\begin{align}
\frac{d}{dt} \hat{\rho}(t) = -\frac{i}{\hbar} [H_{\rm I}(t), \hat{\rho}(t)] , 
\label{LiouvilleEq1}
\end{align}
where the time dependence of the interaction Hamiltonian is obtained by replacing $b_\alpha^\dag \rightarrow e^{i\omega_\alpha t} b_\alpha^\dag$ and $\tilde{c}^\dagger_n \rightarrow e^{i n\Omega t} \tilde{c}^\dagger_n$. 
Equation~\eqref{LiouvilleEq1} can be rewritten in the following integral form:
\begin{align}
\hat{\rho}(t) = \hat{\rho}(0) -\frac{i}{\hbar} \int^{t}_{0} d\tau [H_{\rm I}(\tau), \hat{\rho}(\tau)] ,
\label{LiouvilleEq1_integral}
\end{align}
which can be plugged back into Eq.~\eqref{LiouvilleEq1}:
\begin{align}
\frac{d}{dt} \hat{\rho}(t) 
= &-\frac{i}{\hbar} [H_{\rm I}(t), \hat{\rho}(0)] 
\nonumber \\
&-\frac{1}{\hbar^2} \int_{0}^t d\tau \left[H_{\rm I}(t), [H_{\rm I}(\tau), \hat{\rho}(\tau)] \right], 
\label{LiouvilleEq2}
\end{align}

Now, the reduced density matrix for the system can be obtained by tracing out the bath degrees of freedom from the total density matrix:
\begin{align}
\hat{\rho}_{\rm S} = \mathrm{tr}_{\rm B} \hat{\rho},
\end{align} 
whose dynamics is governed by the corresponding, reduced Liouville equation:
\begin{align}
\frac{d}{dt} \hat{\rho}_{\rm S}(t)
& = - \frac{1}{\hbar^2} \int_{0}^t d\tau \mathrm{tr}_{\rm B} \left[ H_{\rm I}(t), [H_{\rm I}(\tau), \hat{\rho}(\tau)] \right] , 
\label{LiouvilleEq_reduced}
\end{align}
where we set $\mathrm{tr}_{\rm B} [H_{\rm I}(t), \hat{\rho}(0)] = 0$ by assuming that an appropriate initial condition can be imposed on the density matrix. 
Note that Eq.~\eqref{LiouvilleEq_reduced} takes the same form even in the Schr\"{o}dinger picture since 
$[H_{\rm S} + H_{\rm B}, \hat{\rho}_{\rm S}(t)] = 0$.

In what follows, we make various approximations to recast Eq.~\eqref{LiouvilleEq_reduced} into the Lindblad equation, which describes the non-unitary evolution of the reduced density matrix~\cite{Davies74, Lindblad76}.

(i) {\it Born approximation}: 
The system-bath coupling is taken to be sufficiently weak so that the feedback of the system on the bath is negligible. 
As mentioned previously, this amounts to the approximation that the total density matrix is given by the direct product between that of the system and bath: 
$\hat{\rho}(\tau) \simeq \hat{\rho}_{\rm S}(\tau) \otimes \hat{\rho}_{\rm B}$.

(ii) {\it Markov approximation}: 
It is assumed that the dynamics of the system forgets its past history rapidly.
Specifically, this means that $\hat{\rho}_{\rm S}(\tau)$ in Eq.~\eqref{LiouvilleEq_reduced} is replaced by $\hat{\rho}_{\rm S}(t)$, which can be taken out of the integral.

(iii) {\it Born-Markov approximation}: 
In this approximation, the upper limit of the integral in Eq.~\eqref{LiouvilleEq_reduced} is extended to infinity. 
To appreciate the physical meaning of this approximation, it is convenient to consider a specific model for the system-bath coupling, which is quantified via the spectral density:
\begin{align}
J_\textrm{spec}(\omega) = \frac{\pi}{2} \sum_{\alpha=1}^N \frac{C_\alpha^2}{m_\alpha \omega_\alpha} \delta(\omega - \omega_\alpha).
\end{align}
For a proper description of the irreversible process in open systems, the eigenspectrum, $\hbar\omega_\alpha$, needs to be dense so that $J_\textrm{spec}(\omega)$ forms a continuous function~\cite{Leggett87}.

A particularly important case is obtained in the case of the ohmic dissipation, for which the spectral density is given as follows:
\begin{align}
J_\textrm{spec}(\omega) 
= \frac{M \eta \omega}{1 + (\omega / \omega_c)^2},
\end{align} 
where $M$ is the electron mass, $\eta$ is the damping constant, and $\omega_c$ is the high-frequency cutoff. 
One of the most crucial reasons why the ohmic dissipation is important is that the equation of motion reduces to the usual Langevin equation in the semiclassical limit~\cite{Leggett87}.    
Moreover, in this case, it is shown~\cite{Breuer02} that the bath correlation time is given by $\tau_{\rm B} \sim \max\{ \omega_c^{-1}, \hbar / k_{\rm B} T \}$, while the system relaxation time is given by $\tau_{\rm R} \sim \eta^{-1}$.
The Born-Markov approximation becomes valid if $\tau_{\rm B} \ll \tau_{\rm R}$;  
in other words, $\hbar\eta \ll \min\{ \hbar\omega_c, k_{\rm B} T \}$.

Considering the validity regime of the Born-Markov approximation, from now on, we focus on the situation where the damping constant is sufficiently small compared to the high-frequency cutoff, $\omega_c$, and the thermal-fluctuation frequency scale, $k_{\rm B} T/\hbar$.
Strictly speaking, this means that we cannot consider the zero-temperature limit by using the Markovian quantum master equation, which is to be derived with help of the Born-Markov approximation. 
It is important, however, to note that our results remain practically the same even if the temperature is raised slightly.  
For convenience, we set the temperature to be infinitesimally small, or practically zero in this work with a cautionary reminder that the damping constant should be even smaller than the temperature scale.

(iv) {\it Rotating wave approximation}: 
In this approximation, all rapidly oscillating terms proportional to $e^{\pm i\Omega t}$ are ignored in Eq.~\eqref{LiouvilleEq_reduced} under the assumption that the typical time scale for the system evolution, $\tau_{\rm S} \sim \Omega^{-1}$, is sufficiently small compared to $\tau_{\rm R} \sim \eta^{-1}$; in other words, $\eta \ll \Omega$. 
Again, considering the validity regime of the rotating wave approximation, we now focus on the situation where the damping constant is sufficiently small compared to the Bloch oscillation frequency, $\Omega$.

Applying the approximation (i)--(iv) to Eq.~\eqref{LiouvilleEq_reduced}, we finally arrive at the Markovian quantum master, or the Lindblad equation:
\begin{align}
\frac{d}{dt} \hat{\rho}_{\rm S}(t) = -\frac{i}{\hbar} [ H_\mathrm{Lamb}, \hat{\rho}_{\rm S}(t) ] + \mathcal{L}[\hat{\rho}_{\rm S}(t)], 
\label{LindbladEq}
\end{align}
where the Lamb-shift Hamiltonian (describing the energy renormalization) is defined as follows:
\begin{align}
H_\mathrm{Lamb} = \hbar\chi_{+} \sigma_{-} \sigma_{+} + \hbar\chi_{-} \sigma_{+} \sigma_{-} + \hbar\chi_{0} \sigma_{0}^2,
\end{align}
and the Lindbladian (describing the energy dissipation as well as decoherence) is defined as follows:
\begin{align}
\mathcal{L}[\hat{\rho}_{\rm S}(t)]
&= \gamma_{+} \left[ \sigma_{+} \hat{\rho}_{\rm S}(t) \sigma_{-} - \frac{1}{2} \{ \sigma_{-} \sigma_{+}, \hat{\rho}_{\rm S}(t) \} \right] \nonumber\\
&+ \gamma_{-} \left[ \sigma_{-} \hat{\rho}_{\rm S}(t) \sigma_{+} - \frac{1}{2} \{ \sigma_{+} \sigma_{-}, \hat{\rho}_{\rm S}(t) \} \right] \nonumber\\
&+ \gamma_{0} \left[ \sigma_{0} \hat{\rho}_{\rm S}(t) \sigma_{0} - \frac{1}{2} \{ \sigma_{0}^2, \hat{\rho}_{\rm S}(t) \} \right],
\end{align}
where
\begin{align}
\sigma_{\pm}
&= \sum_{n=-\infty}^\infty   \tilde{c}^\dagger_{n\pm1} \tilde{c}_n ,
\\
\sigma_{0}
&= \sum_{n=-\infty}^\infty n \tilde{c}^\dagger_n \tilde{c}_n ,
\end{align}
and 
\begin{align}
\hbar\chi_\pm 
& = \left(\frac{\zeta}{\hbar\Omega}\right)^2 \frac{a^2}{4\pi} 
\mathcal{P} \int_0^\infty d\omega J_\textrm{spec}(\omega) 
\nonumber \\
& ~~~\times \left[ \frac{f_\mathrm{BE}(\omega)}{\omega \mp \Omega} - \frac{1 + f_\mathrm{BE}(\omega)}{\omega \pm \Omega} \right],
\\
\hbar\chi_0 
& = - \frac{\zeta}{\hbar\Omega} \frac{a^2}{2\pi} \mathcal{P} \int_0^\infty d\omega \frac{J_\textrm{spec}(\omega)}{\omega},
\\
\hbar\gamma_{+}
& = \left(\frac{\zeta}{\hbar\Omega}\right)^2 \frac{a^2}{2} J_\textrm{spec}(\Omega) f_\mathrm{BE}(\Omega) , \\
\hbar\gamma_{-}
& =  \left(\frac{\zeta}{\hbar\Omega}\right)^2 \frac{a^2}{2} J_\textrm{spec}(\Omega) \left[ 1+f_\mathrm{BE}(\Omega) \right] ,
\nonumber \\
& = \hbar\gamma_{+} e^{\beta\hbar\Omega}, \\
\hbar\gamma_0
& = \frac{\zeta}{\hbar\Omega} a^2 J_\textrm{spec}(\omega) \left[1 + 2f_\mathrm{BE}(\omega)\right] \Big|_{\omega \rightarrow 0},
\end{align}
with $f_\mathrm{BE}(\omega) = 1 / (e^{\beta\hbar\omega} - 1)$ being the Bose-Einstein distribution function.
Note that, as mentioned before, $\zeta$ has an implicit dependence on ${\bf k}_\perp$, which is dropped here for the sake of convenience.
Restoring the ${\bf k}_\perp$ dependence simply amounts to taking the summation over ${\bf k}_\perp$ in the above equations for $\chi_{\pm,0}$ and $\gamma_{\pm,0}$.

\subsubsection{Thermalization of the WSL eigenstates via the canonical ensemble}
\label{sec:canonical_ensemble}

Now, we investigate the time evolution of a diagonal component of the density matrix, $P_n(t) \equiv \langle n | \hat{\rho}_{\rm S}(t) | n \rangle$, which is the probability that a given (single) electron occupies the $n$-th WSL eigenstate. 
The dynamics of $P_n(t)$ is described by the Pauli master equation:
\begin{align}
\frac{d}{dt} P_n(t) 
& = \gamma_{-} P_{n+1}(t) + \gamma_{+} P_{n-1}(t) - (\gamma_{-} + \gamma_{+}) P_n(t), 
\label{PauliMasterEq}
\end{align}
which is derived from Eq.~\eqref{LindbladEq} by taking the $n$-th diagonal component. 

Under the assumption that $P_n(t)$ does not change in time in the long-time limit, i.~e., $dP_n/dt |_{t\rightarrow\infty} = 0$, the steady state satisfies the following equation: 
\begin{align}
\gamma_{-} (P_{n+1}-P_n) = \gamma_{+} (P_n-P_{n-1}) ,
\label{eq:gen_sol}
\end{align}
which has two solutions.
First, this equation is solved by a solution satisfying the so-called condition of detailed balance:
\begin{align}
\frac{P_{n+1}}{P_{n}} = \frac{P_{n}}{P_{n-1}} =\frac{\gamma_{+}}{\gamma_{-}} = e^{-\beta\hbar\Omega} ,
\label{eq:detailed_balance}
\end{align}
which shows that the occupation probability is governed by the standard canonical ensemble. 
The condition of detailed balance describes the system in the equilibrium ground state, where the WSL eigenstates would be filled with electrons from the low to high energy until the energy reaches the (global) chemical potential energy, leading to inhomogeneous charge density. 
This is basically the {\it electromagnetic shielding}.

As a second solution, we are interested in the alternative nonequilibrium steady state, where the charge density is spatially uniform.
To investigate under which condition such a nonequilibrium steady state can be obtained, we rewrite Eq.~\eqref{eq:detailed_balance} as follows:
\begin{align}
\frac{P_{n+1}}{P_{n}} = \frac{e^{-\beta\epsilon_{n+1}}}{e^{-\beta\epsilon_{n}}} ,
\end{align}
where $\epsilon_n= n\hbar\Omega$ is the energy eigenvalue of the $n$-th WSL eigenstate. 
Considering a generalization to the grand canonical ensemble with many electrons, we introduce the $n$-dependent chemical potential energy, $\mu_n$, as follows:
\begin{align}
\frac{\tilde{P}_{n+1}}{\tilde{P}_{n}} = \frac{e^{-\beta(\epsilon_{n+1}-\mu_{n+1})}}{e^{-\beta(\epsilon_n-\mu_n)}} .
\end{align}
By requiring that the charge density is spatially uniform, i.~e., $\tilde{P}_{n+1}/\tilde{P}_{n}=1$, we arrive at the conclusion that 
\begin{align}
e^{-\beta(\mu_{n+1}-\mu_n)} = e^{-\beta\hbar\Omega} ,
\end{align}
which is equivalent to the condition that $\mu_{n+1}-\mu_n = \hbar\Omega$.
It is important to note that the uniform state satisfying $\tilde{P}_{n+1}/\tilde{P}_{n}=1$ is indeed the second solution of Eq.~\eqref{eq:gen_sol} in addition to the detailed-balance solution.
The charge density becomes half filling if the chemical potential energy is set as $\mu_n=n\hbar\Omega$.

In summary, each WSL eigenstate is individually thermalized via the standard canonical ensemble with its own shifted chemical potential energy.
Combined with the exclusion principle between electrons, this eventually leads to the WSL-wise thermalization for many-electron systems, which is implemented via assigning the Fermi-Dirac distribution to each WSL eigenstate with the shifted chemical potential energy.  
See the following section for details.

\subsection{Conversion from the static scalar to the dynamical vector potential gauge}
\label{sec:conversion}

The noninteracting lesser Green's function, $g^{<}_\mathbf{k}$, is necessary as an input to the Keldysh-Floquet DMFT self-consistency loop.
In this section, we determine $g^{<}_\mathbf{k}$ by converting the above-obtained WSL-wise thermalization scheme from the static scalar to the dynamical vector potential gauge via the gauge invariance principle. 
To this end, let us examine what functional form the local occupation number should take in the static scalar potential gauge, if each WSL eigenstate is individually thermalized via the standard canonical ensemble. 
With the local DOS given by Eq.~\eqref{LDOS_stat}:
\begin{align} 
\rho_{\rm loc}(\omega)
=\sum_l e^{-\frac{1}{2}(t^*/\hbar\Omega)^2} I_l ((t^*/\hbar\Omega)^2/2 ) \rho_{{\rm lad}, l}(\omega) ,
\end{align}  
it is natural to infer that the local occupation number should take the following form:
\begin{align} 
N_{\rm loc}(\omega)
= &~\sum_l e^{-\frac{1}{2}(t^*/\hbar\Omega)^2} I_l ((t^*/\hbar\Omega)^2/2 ) \rho_{{\rm lad}, l}(\omega) 
\nonumber \\
&\times f_{\rm FD}(\omega-l\Omega),
\label{eq:N_loc_non-int}
\end{align} 
where $f_{\rm FD}(\omega)=1/(e^{\beta\hbar\omega}+1)$ is the Fermi-Dirac distribution function. 
The rationale behind Eq.~\eqref{eq:N_loc_non-int} is that the spectral weight of each WSL eigenstate, say, $e^{-\frac{1}{2}(t^*/\hbar\Omega)^2} I_l ((t^*/\hbar\Omega)^2/2 ) \rho_{{\rm lad}, l}(\omega)$ for the $l$-th WSL eigenstate, is individually thermalized by its own Fermi-Dirac distribution function, $f_{\rm FD}(\omega-l\Omega)$. 
As mentioned in the preceding section, the $l$-th WSL eigenstate is accompanied with its own shifted chemical potential, $l\hbar\Omega$.

Now, it is reminded that the local occupation number is a physical observable so that it should be gauge invariant. 
By using this gauge invariance principle, one can now determine $g^<_{\bf k}(\omega)$ in the dynamical vector potential gauge.
Specifically, in the dynamical vector potential gauge, one can obtain exactly the same local occupation number as given in Eq.~\eqref{eq:N_loc_non-int} by taking the imaginary part of the following lesser Green's function:
\begin{align}
N_{\rm loc}(\omega+n\Omega) = \frac{1}{2\pi} \mathrm{Im} \sum_{\bf k} [g^<_{\bf k}(\omega)]_{nn} ,
\end{align}
where
\begin{align}
g_{\mathbf{k}}^<(\omega) 
&= \mathcal{U}_\mathbf{k} \cdot  g_\mathrm{lad}^<(\omega) \cdot  \mathcal{U}_\mathbf{k}^\dag
\nonumber \\
&=-\mathcal{U}_\mathbf{k} \cdot  \left\{g_\mathrm{lad}^r(\omega) \mathcal{F}(\omega) - \mathcal{F}(\omega) [g_\mathrm{lad}^{r}(\omega)]^\dagger \right\} \cdot  \mathcal{U}_\mathbf{k}^\dag,
\label{LesserWSL}
\end{align}
where $\mathcal{F}_{mn}(\omega)=f_\mathrm{FD}(\omega+n\Omega) \delta_{mn}$with $-\Omega/2 < \omega \leq \Omega/2$. 
The above form can be deduced from the fact that (i)  the WSL Green's function, $g^r_{\rm lad}$, is formally identical to that of an energy eigenmode and, 
(ii) if so, $g_\mathrm{lad}^<(\omega) = -g_\mathrm{lad}^r(\omega) \mathcal{F}(\omega) + \mathcal{F}(\omega) [g_\mathrm{lad}^{r}(\omega)]^\dagger$ according to the standard fluctuation-dissipation theorem. 
The use of the fluctuation-dissipation theorem in the dynamical vector gauge is consistent with the assumption that the WSL eigenstates are thermalized via the standard canonical ensemble in the static scalar potential gauge. 
It is important to note that, with proper definition, the (total) energy is still conserved in the dynamical vector potential gauge, while the expectation value of the Hamiltonian is not.  
See Appendix~\ref{appen:energy_conservation} for details.

The so-obtained local distribution function, $f_{\rm loc}(\omega) = N_{\rm loc}(\omega)/\rho_{\rm loc}(\omega)$, for the noninteracting system is shown in Fig.~\ref{Fig2_SC_loop}~(c) by using the color scale as a function of energy for each lattice site, which is represented as a quantum well for the sake of clarity.  
As mentioned in Sec.~\ref{sec:master_equation}, the actual temperature inside the Fermi-Dirac distribution function, $f_\mathrm{FD}(\omega)$, (which is dictated by the thermal bath) is set to be infinitesimally small, or practically zero throughout this work with a cautionary reminder that the damping constant is even smaller than the temperature scale.
It is important to note that $f_{\rm loc}(\omega)$ reduces to $f_\mathrm{FD}(\omega)$ in the low-field limit.

It is interesting to mention that, in the fermion-bath model~\cite{Han13, Tsuji09, Amaricci12, Aron12a, Aron12b}, the lesser Green's function takes the form of  
$g_{\mathbf{k}}^<(\omega) 
= g_{\mathbf{k}}^r(\omega) \cdot \Sigma_\mathrm{diss}^<(\omega) \cdot [g_{\mathbf{k}}^r(\omega)]^\dag$,
where the self-energy, which is responsible for the energy dissipation, is given by
$\Sigma_\mathrm{diss}^<(\omega) = i\Gamma \mathcal{F}(\omega)$. 
This form for the lesser Green's function is definitely different from Eq.~\eqref{LesserWSL} obtained in the WSL-wise thermalization scheme, resulting in different local distribution functions, while both share the same retarded part. 
Intriguingly, the WSL-wise thermalization can be formally reproduced in this format if $\Sigma_\mathrm{diss}^<(\omega)$ were replaced by $\mathcal{U}_\mathbf{k} \cdot  \Sigma_\mathrm{diss}^<(\omega) \cdot  \mathcal{U}_\mathbf{k}^\dag$.

\subsection{Self-consistency loop for the Keldysh-Floquet DMFT}
\label{sec:KeldyshDyson}

It is mentioned in the beginning of Sec.~\ref{sec:DMFT} that the self-consistency loop for the Keldysh-Floquet DMFT can be constructed concretely, based on the Keldysh-contour expansion. 
In this section, we provide the detailed Floquet-matrix form of the Keldysh-Dyson equation, which is embedded in the Keldysh-Floquet DMFT self-consistency loop. 
Note that similar forms can be found in previous works~\cite{Tsuji08, Aron12a, Aron12b, Tsuji09}, while organized differently.

\subsubsection{Lattice Green's function}

We begin by writing the Dyson equation for the retarded lattice Green's function as follows:
\begin{align}
[G_\mathbf{k}^r(\omega)]^{-1} = [g_\mathbf{k}^r(\omega)]^{-1} - \Sigma^r(\omega) ,
\label{DysonEq}
\end{align}
where, according to the usual DMFT philosophy, it is assumed that the self-energy is independent of the momentum, ${\bf k}$. 
Furthermore, the self-energy is assumed to be diagonal in the Floquet-matrix form since all physical observables in the steady state should be independent of the change of the time origin, i.~e., the time translation. 
Note that the invariance with respect to the time translation is equivalent to the diagonal condition in the Floquet-matrix form.

Interestingly, the time translation can be viewed as a gauge transformation~\cite{Tsuji08}.
In this view, it is natural to infer that all local Green's functions, which are gauge invariant~\cite{Tsuji08}, should be diagonal in the Floquet-matrix form. 
Similarly, being a local observable, the self-energy should be also diagonal.
At this point, it is important to note that all physical quantities in our Keldysh-Floquet DMFT, which are local themselves, are related with each other via local equations and therefore all obtained results are guaranteed to be gauge invariant.

Concretely, each Floquet-matrix element of Eq.~\eqref{DysonEq} can be written as follows:
\begin{align}
&[G^r]^{-1}_{mn}(\zeta_{\bf k},\theta_{\bf k},\omega)
\nonumber \\
&= e^{i(m - n)\theta_{\bf k}} \Big\{
\left[ \hbar(\omega + n\Omega) + i\Gamma/2 - \Sigma^r_{nn}(\omega) \right] \delta_{mn} 
\nonumber \\
&- (\zeta_{\bf k}/2) (\delta_{m, n+1} + \delta_{m, n-1})
\Big\} ,
\end{align}
which, being a tridiagonal matrix, can be inverted analytically~\cite{Huang97}.
Specifically, the diagonal elements have the structure of a continued fraction:
\begin{align}
&G_{nn}^r(\zeta_{\bf k},\theta_{\bf k},\omega) 
\nonumber \\
&= \Big\{
\hbar(\omega + n\Omega) + i\Gamma/2 - \Sigma^r_{nn}(\omega) 
\nonumber \\
&- (\zeta_{\bf k}/2)^2 
\left[P_{n-1}^{(-)}(\zeta_{\bf k},\omega) + P_{n+1}^{(+)}(\zeta_{\bf k},\omega)\right]
\Big\}^{-1}, 
\label{InvRetGreen_diag}
\end{align}
where $P^{(\pm)}_n$ is determined recursively as follows:
\begin{align}
P_{n}^{(\pm)}(\zeta_{\bf k},\omega) 
&= \Big[
\hbar(\omega +n\Omega) + i\Gamma/2 - \Sigma^r_{nn}(\omega)  
\nonumber \\
&- (\zeta_{\bf k}/2)^2 P^{(\pm)}_{n \pm 1} (\zeta_{\bf k},\omega)
\Big]^{-1} ,
\end{align}
which satisfies the boundary condition that $P^{(\pm)}_{\pm N_{\rm c}} =0$ for sufficiently large $N_{\rm c}$, which denotes the cutoff index for high-frequency Floquet modes. 
Meanwhile, the off-diagonal elements can be determined from the diagonal counterparts as follows:
\begin{align}
&G_{mn}^r(\zeta_{\bf k},\theta_{\bf k},\omega) 
\nonumber \\
&=
\begin{cases}
e^{i\theta_{\bf k}} \frac{\zeta_{\bf k}}{2} P_m^{(+)}(\zeta_{\bf k},\omega) G_{m-1,n}^r(\zeta_{\bf k},\theta_{\bf k},\omega) 
& \textrm{for $m>n$,}
\\
\\
e^{-i\theta_{\bf k}} \frac{\zeta_{\bf k}}{2} P_n^{(+)}(\zeta_{\bf k},\omega) G_{m,n-1}^r(\zeta_{\bf k},\theta_{\bf k},\omega) 
& \textrm{for $m<n$}.
\end{cases}
\label{InvRetGreen_off} 
\end{align}

Next, let us write the Keldysh equation for the lesser Green's function, which can be reduced as follows:
\begin{widetext}
\begin{align}
G_\mathbf{k}^<(\omega) 
&= \big[1 + G_\mathbf{k}^r(\omega) \cdot \Sigma^r(\omega) \big] \cdot g_\mathbf{k}^<(\omega) \cdot \big[1 + \Sigma^a(\omega) \cdot G_\mathbf{k}^a(\omega) \big] 
+ G_\mathbf{k}^r(\omega) \cdot \Sigma^<(\omega) \cdot G_\mathbf{k}^a(\omega) 
\nonumber\\
&= G_\mathbf{k}^r(\omega) \cdot [g_\mathbf{k}^r(\omega)]^{-1} \cdot g_\mathbf{k}^<(\omega) \cdot [g_\mathbf{k}^a(\omega)]^{-1} \cdot G_\mathbf{k}^a(\omega) 
+ G_\mathbf{k}^r(\omega) \cdot \Sigma^<(\omega) \cdot G_\mathbf{k}^a(\omega) 
\nonumber\\
&= G_\mathbf{k}^r(\omega) \cdot \big\{ \mathcal{U}_\mathbf{k} \cdot [g_\mathrm{lad}^r(\omega)]^{-1} \cdot g_\mathrm{lad}^<(\omega) \cdot [g_\mathrm{lad}^a(\omega)]^{-1} \cdot \mathcal{U}_\mathbf{k}^\dag  
+ \Sigma^<(\omega) \big\} \cdot G_\mathbf{k}^a(\omega) 
\nonumber\\
&= G_\mathbf{k}^r(\omega) \cdot \big[ \mathcal{U}_\mathbf{k} \cdot i\Gamma {\cal F}(\omega) \cdot \mathcal{U}_\mathbf{k}^\dag + \Sigma^<(\omega) \big] \cdot G_\mathbf{k}^a(\omega),
\label{KeldyshEq_main}
\end{align}
\end{widetext}
where the second line is obtained by Eq.~\eqref{DysonEq} and the third line is obtained by 
Eqs.~\eqref{RetWSL} and \eqref{LesserWSL}. 
Note that the last line is nothing but the fluctuation-dissipation theorem for $g_\mathrm{lad}^<$.
Concretely, Eq.~\eqref{KeldyshEq_main} can be rewritten for each element of the Floquet matrix as follows:
\begin{widetext}
\begin{align}
G_{mn}^<(\zeta_\mathbf{k},\theta_\mathbf{k},\omega) 
& = e^{i(m-n)\theta_\mathbf{k}} \bigg[ \sum_{l,p,q=-N_c}^{N_c}  G_{mp}^r(\zeta_\mathbf{k},\theta_\mathbf{k}=0,\omega) J_{p-l}(\zeta_\mathbf{k}/\hbar\Omega) 
i \Gamma f_\mathrm{FD}(\omega + l\Omega) J_{q-l}(\zeta_\mathbf{k}/\hbar\Omega)  [G_{nq}^r(\zeta_\mathbf{k},\theta_\mathbf{k}=0,\omega)]^* 
\nonumber\\
& + \sum_{l=-N_c}^{N_c}  G_{ml}^r(\zeta_\mathbf{k},\theta_\mathbf{k}=0,\omega) \Sigma_{ll}^<(\omega)  [G_{nl}^r(\zeta_\mathbf{k},\theta_\mathbf{k}=0,\omega)]^*  \bigg] ,
\end{align}
\end{widetext}
where, for convenience, the $\theta$-dependent exponential factor is taken out explicitly to the front of the expression by using the fact that $G_{mn}^{r}(\zeta,\theta,\omega)=e^{i(m-n)\theta} G_{nn}^{r}(\zeta,\theta=0,\omega)$.

\subsubsection{Local Green's function}

The local Green's function can be obtained by summing the lattice Green's function over all momenta:
\begin{align}
G_{mn}^{r,<}(\omega) 
& = \sum_\mathbf{k}(G_\mathbf{k}^{r,<})_{mn}(\omega) \nonumber\\
& = \int_0^{2\pi} d\theta \int_0^\infty d\zeta~\zeta~
\rho_\mathrm{non}(\zeta) G_{mn}^{r,<}(\zeta,\theta,\omega) \nonumber\\
& = 2\pi \delta_{mn} \int_0^\infty d\zeta~\zeta~\rho_\mathrm{non}(\zeta)G_{nn}^{r,<}(\zeta,\theta=0,\omega),
\end{align}
where the last line is obtained since $G_{mn}^{r,<}(\zeta,\theta,\omega)=e^{i(m-n)\theta} G_{nn}^{r,<}(\zeta,\theta=0,\omega)$, whose $\theta$-integration, in turn, gives rise to $\delta_{mn}$. 
Note that this result is consistent with the requirement that the local Green's function should be gauge invariant and therefore diagonal.

\subsubsection{Weiss function}

The Weiss function is the ``noninteracting'' local Green's function of the effective impurity model, onto which the original lattice model is mapped under the DMFT self-consistency condition.
The retarded Weiss function is related with the fully-interacting retarded local Green's function via the Dyson equation,
\begin{align}
[G^r(\omega)]^{-1} 
= [\mathcal{G}_0^r(\omega)]^{-1} - \Sigma_U^r(\omega),
\label{WeissDysonEq}
\end{align}
which involves only the self-energy contributed by the electron-electron interaction, $\Sigma^r_U$.  
The self-energy contribution from the electron-impurity interaction, $\Sigma^r_\mathrm{imp}$, is excluded here since the Weiss function is interpreted as the local Green's function in the absence of the on-site electron-electron interaction.
Note that the total self-energy, $\Sigma=\Sigma_U+\Sigma_{\rm imp}$, is used in the Keldysh-Dyson equation for the lattice Green's function in Eqs.~\eqref{DysonEq} and \eqref{KeldyshEq_main}. 
Now, since both the local Green's function and self-energy are diagonal Floquet matrices, Eq.~\eqref{WeissDysonEq} can be inverted algebraically as follows:
\begin{align}
\mathcal{G}_0^r(\omega) 
= \{ [G^r(\omega)]^{-1} + \Sigma_U^r(\omega) \}^{-1} . 
\end{align}

Next, the lesser Weiss function can be determined via the Keldysh equation:
\begin{align}
G^<(\omega) 
&=\left[1 + G^r(\omega)\Sigma_U^r(\omega)\right]
\mathcal{G}_0^<(\omega)
\left[1 + \Sigma_U^a(\omega) G^a(\omega)\right] 
\nonumber\\
&+ G^r(\omega)\Sigma_U^<(\omega) G^a(\omega) ,
\end{align}
which can be eventually rearranged as follows:
\begin{align}
\mathcal{G}_0^<(\omega) 
= |\mathcal{G}_0^r(\omega)|^2 \left[ G^<(\omega) / |G^r(\omega)|^2 - \Sigma_U^<(\omega) \right] .
\end{align}

\subsubsection{Self-energy}

As mentioned in Sec.~\ref{sec:Hamiltonian}, we consider two interaction terms in the Hamiltonian with one being the electron-electron interaction, $U$, and the other being the electron-impurity interaction, $V$, which generate the self-energies, $\Sigma_U$ and $\Sigma_\mathrm{imp}$, respectively.

First, $\Sigma_U$ is computed via the iterated perturbation theory (IPT) as an impurity solver. 
The IPT self-energy is written in real time as follows:
\begin{align}
\hbar \Sigma_U^t(t,t') = U^2[\hbar \mathcal{G}_0^t(t,t')]^2 \hbar \mathcal{G}_0^t(t',t),
\end{align}
which is analytically continued to the Keldysh contour via the Langreth theorem~\cite{Haug08} as follows:
\begin{align}
\hbar \Sigma_U^\lessgtr(t,t') = U^2[\hbar \mathcal{G}_0^\lessgtr(t,t')]^2 \hbar \mathcal{G}_0^\gtrless(t',t) ,
\end{align}
which further reduces to the following form under the assumption that the self-energy depends only on the relative time in the steady state:
\begin{align}
\hbar \Sigma_U^\lessgtr(t_\mathrm{rel})
= U^2 [\hbar \mathcal{G}_0^\lessgtr(t_\mathrm{rel})]^2 \hbar \mathcal{G}_0^\gtrless(-t_\mathrm{rel}) .
\label{LesIPTtime}
\end{align}
The retarded self-energy is connected with the lesser and greater counterpart via 
\begin{align}
\Sigma_U^r(t_\mathrm{rel}) = \theta(t_\mathrm{rel})[\Sigma_U^>(t_\mathrm{rel}) - \Sigma_U^<(t_\mathrm{rel})]. 
\label{RetIPTtime}
\end{align}
Finally, Eqs.~\eqref{LesIPTtime} and \eqref{RetIPTtime} are converted to the Floquet-matrix form by
\begin{align}
\left[ \Sigma^{r,<}_U (\omega) \right]_{mn} = \delta_{mn} \int d t_\textrm{rel} e^{i (\omega+n\Omega) t_\textrm{rel}} \Sigma^{r,<}_U (t_\textrm{rel}).
\end{align}

Second, $\Sigma_\mathrm{imp}$ is computed via the self-consistent Born approximation (SCBA) for the on-site electron-impurity interaction, $V$, in which case $\Sigma_\mathrm{imp}$ becomes a local quantity and therefore its computation can be absolved nicely into the DMFT framework. 
Specifically, $\Sigma_\mathrm{imp}$ can be directly connected to the local Green's function as follows:
\begin{align}
\Sigma^{r,<}_\mathrm{imp}(\omega) = n_\textrm{imp} V^2 G^{r,<}(\omega),
\end{align}
with $n_\textrm{imp}$ being the average impurity number per site.

\subsection{Validity of the iterated perturbation theory (IPT)}
\label{sec:IPT}

In equilibrium, the IPT can be regarded as an interpolation scheme connecting between the weak and strong-$U$ limit, where the IPT becomes exact. 
In this section, we show that the same argument can be applied by proving that the IPT is exact in both limits of weak and strong $U$ even within the Keldysh-Floquet DMFT.

First, the IPT is exact in the limit of weak $U$ since $\mathcal{G}_0$ becomes identical to the noninteracting local Green's function, $g=\sum_\mathbf{k} g_\mathbf{k}$, in which case the IPT reduces to a simple second-order perturbation theory. 
It is important to note that, with reformulation via the Floquet representation, the validity of the IPT is extended for arbitrary electric field strength in the weak-$U$ limit.

Second, the exactness of the IPT is proved in the strong-$U$ limit by examining the exact form of the local Green's function:
\begin{align}
G^r_{mn}(\omega) 
= &~ \frac{1}{2}\delta_{mn} \bigg[\frac{1}{\hbar(\omega+n\Omega)+U/2+i\eta} 
\nonumber \\
& + \frac{1}{\hbar(\omega+n\Omega)-U/2+i\eta} \bigg] ,
\label{RetGreenLargeU}
\\ 
G^<_{mn}(\omega) 
= & -2i \delta_{mn} f_\textrm{FD}(\omega+n\Omega) \mathrm{Im} G^r_{nn}(\omega) , 
\label{LesGreenLargeU}
\end{align} 
which are obtained by realizing that the strong-$U$ limit is nothing but taking the hopping amplitude to be zero, i.~e., $t^* \rightarrow 0$, in which case 
(i) all sites become completely isolated, and (ii) the Floquet mode index corresponds simply to the site index.
The form of the retarded Green's function is determined by the fact that, in this limit, the Hubbard bands are simply composed of two completely independent sharp peaks with their centers located at $U/2$ and $-U/2$.  
On the other hand, the form of the lesser counterpart is dictated by the fact that the Hubbard levels (which are energy eigenstates) are thermalized according to the standard canonical ensemble. 

The next step is to deduce the Weiss function as well as the self-energy from the exact retarded Green's function in Eq.~\eqref{RetGreenLargeU}, which can be performed by noting that 
\begin{align}
(G^r)^{-1}_{mn}(\omega)&=\delta_{mn}\left[ \hbar(\omega+n\Omega)+i\eta -\frac{(U/2)^2}{\hbar(\omega+n\Omega)+i\eta} \right]
\nonumber \\
&=(\mathcal{G}^r_0)^{-1}_{mn}(\omega) - (\Sigma_U^r)_{mn}(\omega),
\end{align} 
where
\begin{align}
(\mathcal{G}^r_0)_{mn}(\omega) 
& = \delta_{mn} \frac{1}{\hbar(\omega+n\Omega)+i\eta}, 
\label{RetWeissLargeU}
\\
(\Sigma_U^r)_{mn}(\omega) 
& = \delta_{mn} (U/2)^2 (\mathcal{G}_0^r)_{nn}(\omega) ,
\label{RetSelfLargeU}
\end{align}
which, in turn, dictate the forms of the lesser counterparts:
\begin{align} 
(\mathcal{G}^<_0)_{mn}(\omega) &= -2i \delta_{mn} f_\textrm{FD}(\omega+n\Omega) \mathrm{Im} (\mathcal{G}^r_0)_{nn}(\omega),
\label{LesWeissLargeU}
\\
(\Sigma_U^<)_{mn}(\omega) 
& = \delta_{mn} (U/2)^2 (\mathcal{G}_0^<)_{nn}(\omega) ,
\label{LesSelfLargeU}
\end{align}
where Eq.~\eqref{LesWeissLargeU} is obtained since Eq.~\eqref{RetWeissLargeU} has exactly the same form as the noninteracting  electron propagator, while Eq.~\eqref{LesSelfLargeU} is a consequence of the Langreth rule. 
Now, what needs to be proved is that the above-obtained $\mathcal{G}_0$ and $\Sigma_U$ are connected with each other through the IPT impurity solver.
In other words, Eqs.~\eqref{RetSelfLargeU} and \eqref{LesSelfLargeU} are derived from Eqs.~\eqref{LesIPTtime} and \eqref{RetIPTtime} with $\mathcal{G}_0^{r,<}$ given by Eqs.~\eqref{RetWeissLargeU} and \eqref{LesWeissLargeU}.

To this end, we first perform Fourier transformation of Eq.~\eqref{LesWeissLargeU}:
\begin{align}
\mathcal{G}_0^<(t_\mathrm{rel})
& = \sum_{n=-\infty}^\infty \int_{-\Omega/2}^{\Omega/2} \frac{d\omega}{2\pi} e^{-i (\omega + n\Omega) t_\mathrm{rel}} (\mathcal{G}_0^<)_{nn}(\omega) 
\nonumber \\
&= i f_\mathrm{FD}(0) / \hbar = i / (2\hbar) , 
\label{LesWeissLargeUtime}
\end{align}
whose greater counterpart is similarly obtained as follows:
\begin{align}
\mathcal{G}_0^>(t_\mathrm{rel})
& = \sum_{n=-\infty}^\infty \int_{-\Omega/2}^{\Omega/2} \frac{d\omega}{2\pi} e^{-i (\omega + n\Omega) t_\mathrm{rel}} (\mathcal{G}_0^>)_{nn}(\omega) 
\nonumber \\
&= -i [1 - f_\mathrm{FD}(0)] / \hbar = - i / (2\hbar). 
\label{GrtWeissLargeUtime}
\end{align}
Then, by plugging Eqs.~\eqref{LesWeissLargeUtime} and \eqref{GrtWeissLargeUtime} into Eq.~\eqref{LesIPTtime}, one obtains
\begin{align}
\hbar \Sigma_U^\lessgtr(t_\mathrm{rel})
= \pm i U^2 / 8,
\end{align}
which, according to Eq.~\eqref{RetIPTtime}, gives rise to 
\begin{align}
\hbar \Sigma_U^r (t_\mathrm{rel})
= -i \theta(t_\mathrm{rel}) (U/2)^2 .
\label{RetSelfLargeUtime}
\end{align}
Finally, Fourier transforming Eq.~\eqref{RetSelfLargeUtime} to the frequency domain generates the desired result:
\begin{align}
(\Sigma_U^r)_{nn}(\omega)
&= \int_{-\infty}^\infty dt_\mathrm{rel} e^{i (\omega + n\Omega) t_\mathrm{rel}} \Sigma_U^r (t_\mathrm{rel}) 
\nonumber \\
&= \frac{(U/2)^2}{\hbar(\omega+n\Omega)+i\eta} \nonumber\\
&= (U/2)^2 (\mathcal{G}_0^r)_{nn}(\omega) ,
\end{align}
where we use that
$\int_{-\infty}^\infty \frac{d\omega}{2\pi} e^{-i \omega t_\mathrm{rel}} / (\hbar\omega + i\eta) = -\frac{i}{\hbar} \theta(t_\mathrm{rel})$.
In conclusion, the IPT produces the exact retarded self-energy in the strong-$U$ limit.

The lesser part of the IPT is proved in the sense that, being a consequence of the Langreth rule, Eq.~\eqref{LesSelfLargeU} is  completely dictated by Eq.~\eqref{RetSelfLargeU}. 
To be absolutely sure, we check if the above usage of the Langreth rule is consistent with our assignment for $\mathcal{G}_0^<$ in Eq.~\eqref{LesWeissLargeU}.
To do so, let us examine the Keldysh equation for the lesser local Green's function, $G_{mn}^< = \delta_{mn} G_{nn}^<$:
\begin{widetext}
\begin{align}
G_{nn}^<(\omega) 
& = \left[1 + G_{nn}^r(\omega)(\Sigma_U^r)_{nn}(\omega)\right] (\mathcal{G}_0^<)_{nn}(\omega) \left[1 + (\Sigma_U^a)_{nn}(\omega) G_{nn}^a(\omega)\right] + G_{nn}^r(\omega)(\Sigma_U^<)_{nn}(\omega) G_{nn}^a(\omega) 
\nonumber \\
& = G_{nn}^r(\omega) \left[(G^r)^{-1}_{nn}(\omega) + (\Sigma_U^r)_{nn}(\omega)\right] (\mathcal{G}_0^<)_{nn}(\omega) \left[(G^a)^{-1}_{nn}(\omega) + (\Sigma_U^a)_{nn}(\omega) \right] G_{nn}^a(\omega) + G_{nn}^r(\omega)(\Sigma_U^<)_{nn}(\omega) G_{nn}^a(\omega) 
\nonumber \\
& = G_{nn}^r(\omega) \left[ (\mathcal{G}_0^r)^{-1}_{nn}(\omega)  (\mathcal{G}_0^<)_{nn}(\omega) (\mathcal{G}_0^a)^{-1}_{nn}(\omega) + (\Sigma_U^<)_{nn}(\omega) \right] G_{nn}^a(\omega) 
\nonumber \\
& = \frac{G_{nn}^r(\omega) - G_{nn}^a(\omega)}{(G^a)_{nn}^{-1}(\omega) - (G^r)_{nn}^{-1}(\omega)} 
\left[ \frac{(\mathcal{G}_0^a)^{-1}_{nn}(\omega) -  (\mathcal{G}_0^r)^{-1}_{nn}(\omega)}{(\mathcal{G}_0^r)_{nn}(\omega) - (\mathcal{G}_0^a)_{nn}(\omega)} (\mathcal{G}_0^<)_{nn}(\omega) + (\Sigma_U^<)_{nn}(\omega) \right] ,
\label{Keldysh}
\end{align}
\end{widetext}
where the third line is obtained by using the Dyson equation for the retarded local Green's function, while the fourth line is by the simple fact that both $G_{mn}$ and $(\mathcal{G}_0)_{mn}$ are diagonal matrices so that $(G)^{-1}_{nn}=1/G_{nn}$ and $(\mathcal{G}_0)^{-1}_{nn}=1/(\mathcal{G}_0)_{nn}$.
Then, by rearranging Eq.~\eqref{Keldysh},  $\Sigma_U^<$ is obtained as follows:
\begin{align} 
(\Sigma_U^<)_{nn}(\omega)
&= \left[ (G^a)_{nn}^{-1}(\omega) - (G^r)_{nn}^{-1}(\omega) \right] \frac{G_{nn}^<(\omega)}{2i\mathrm{Im}G_{nn}^r(\omega)} 
\nonumber \\
&~ - \left[ (\mathcal{G}_0^a)^{-1}_{nn}(\omega) 
-  (\mathcal{G}_0^r)^{-1}_{nn}(\omega) \right]  \frac{(\mathcal{G}_0^<)_{nn}(\omega)}{2i\mathrm{Im}(\mathcal{G}_0^r)_{nn}(\omega)} 
\nonumber \\
& = - \big[ (\Sigma_U^r)_{nn}(\omega) - (\Sigma_U^a)_{nn}(\omega) \big] f_\mathrm{FD}(\omega+n\Omega)
\nonumber \\
& = -2i f_\mathrm{FD}(\omega+n\Omega) (U/2)^2 \mathrm{Im} (\mathcal{G}_0^r)_{nn}(\omega) 
\nonumber \\
& = (U/2)^2 (\mathcal{G}_0^<)_{nn}(\omega),
\label{LesSelfLargeU2} 
\end{align}
where the second line is obtained by using Eqs.~\eqref{LesGreenLargeU} and \eqref{LesWeissLargeU} as well as the Dyson equation, while the fourth line is obtained by using Eq.~\eqref{RetSelfLargeU}.
As one can see, Eq.~\eqref{LesSelfLargeU2} is identical to Eq.~\eqref{LesSelfLargeU}, completing the desired proof.

Actually, a slightly more thorough analysis can be performed for the strong-$U$ limit, which is beyond simply taking the hopping amplitude to be zero. 
See Appendix~\ref{appen:IPT} for details.  
In conclusion, the IPT can be regarded as an interpolation scheme connecting between the weak and strong-$U$ limit even within the Keldysh-Floquet DMFT.

\section{Results}
\label{sec:results}

\subsection{Local spectrum}
\label{sec:local_spectrum}

\subsubsection{Emergent Wannier-Stark ladder}
\label{sec:emergentWSL}

\begin{figure}
\centering
\includegraphics[width=0.43\textwidth]
{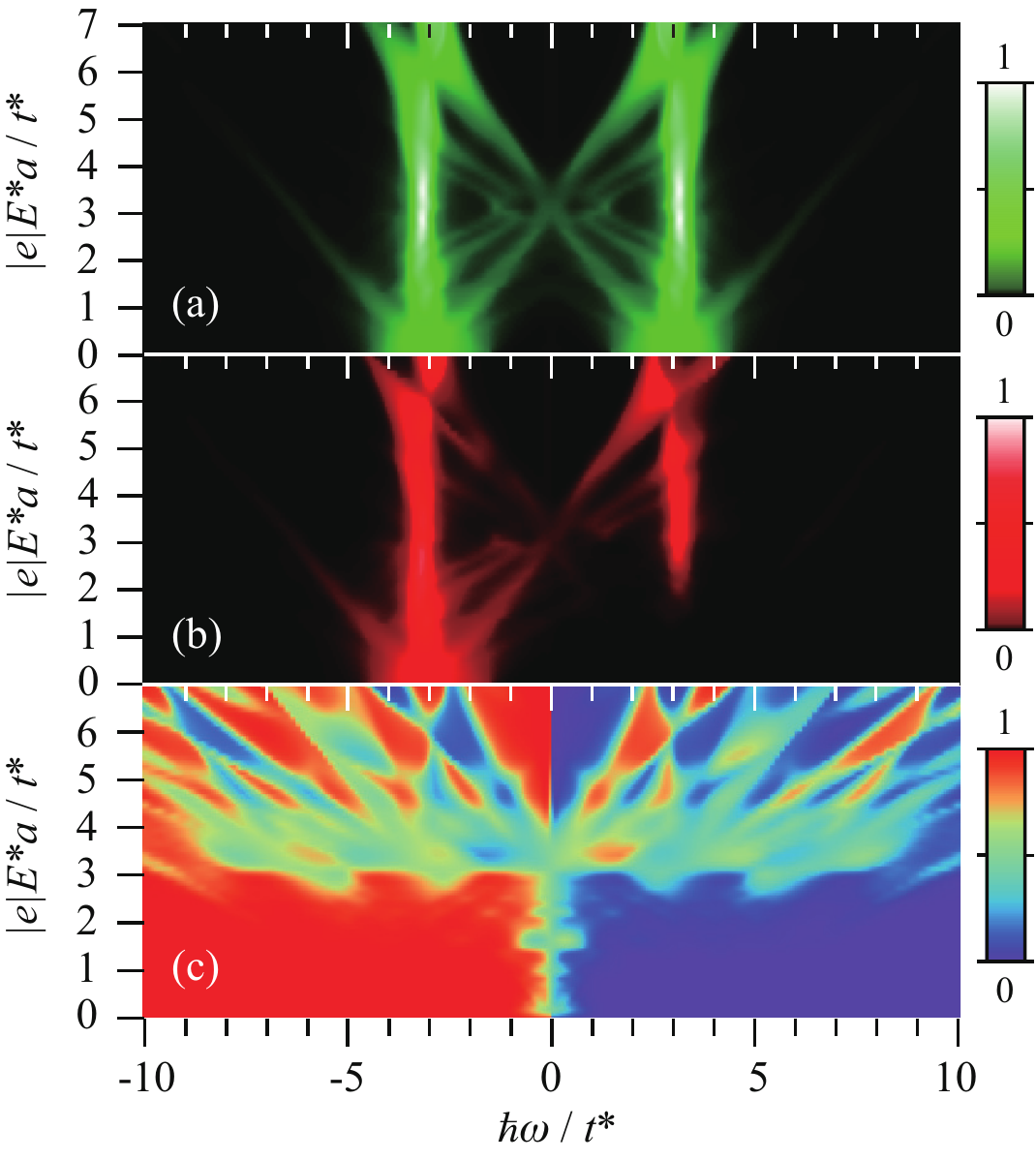} \\
\caption{
Evolution of the local (a) DOS, $\rho_\mathrm{loc}(\omega)$, (b) occupation number, $N_\mathrm{loc}(\omega)$, and (c) distribution function, $f_\mathrm{loc}(\omega)$, as a function of $E^*$ for the electric-field-driven Mott insulator at $U / t^* = 6$, $V_\textrm{imp} / t^* = 0.2$, and $\Gamma/t^*=0.01$.    
}
\label{Fig3_Spectrum}
\end{figure}

Figure~\ref{Fig3_Spectrum} shows the evolution of the local (a) DOS, (b) occupation number, and (c) distribution function as a function of $E^*$ at $U / t^* = 6$ and $V_\textrm{imp} / t^* = 0.2$, where $V_\textrm{imp}=\sqrt{n_\textrm{imp}} V$. 
For convenience, here, we choose a small, but finite level broadening width; $\Gamma/t^*=0.01$. 
It is shown in Fig.~\ref{Fig3_Spectrum}~(a) that the local DOS is reconstructed within the Mott gap at $E^* = E^*_\textrm{mer} \simeq U/2|e|a$, where two of the WSL branches respectively emanating from the upper and lower Hubbard band merge at $\omega=0$. 
Note that the upper and lower Hubbard band generate their own WSL branches, whose positions are respectively described by the formula, $\hbar\omega \simeq n \hbar \Omega \pm U/2$ with $n$ being an integer and $\hbar\Omega=|e|E^* a$. 
The local DOS is maximally reconstructed when the $n=1$ WSL branch from the lower Hubbard band merges with the $n=-1$ counterpart from the upper Hubbard band, i.~e., when $\hbar \Omega \simeq U/2$. 
Interestingly, a similar merging of the $|n|=2$ WSL branches is observed with a hint of the secondary reconstruction of the local DOS around $\hbar\Omega \simeq U/4$.

The reconstruction of the local DOS at $\omega=0$ via merging of the WSL branches is overall consistent with the previous result originally obtained by Joura and collaborators~\cite{Joura08}.
A similar feature has been also observed in the Falikov-Kimball model~\cite{Tsuji08}, where there is an exact solution for the impurity problem within the DMFT even in nonequilibrium. 
It is interesting to note that, while the actual merging was not clearly identified, the existence of the WSL branches emanating from the Hubbard bands was observed in other previous approaches~\cite{Aron12b, Eckstein13} as well.

\begin{figure}
\centering
\includegraphics[width=0.43\textwidth]
{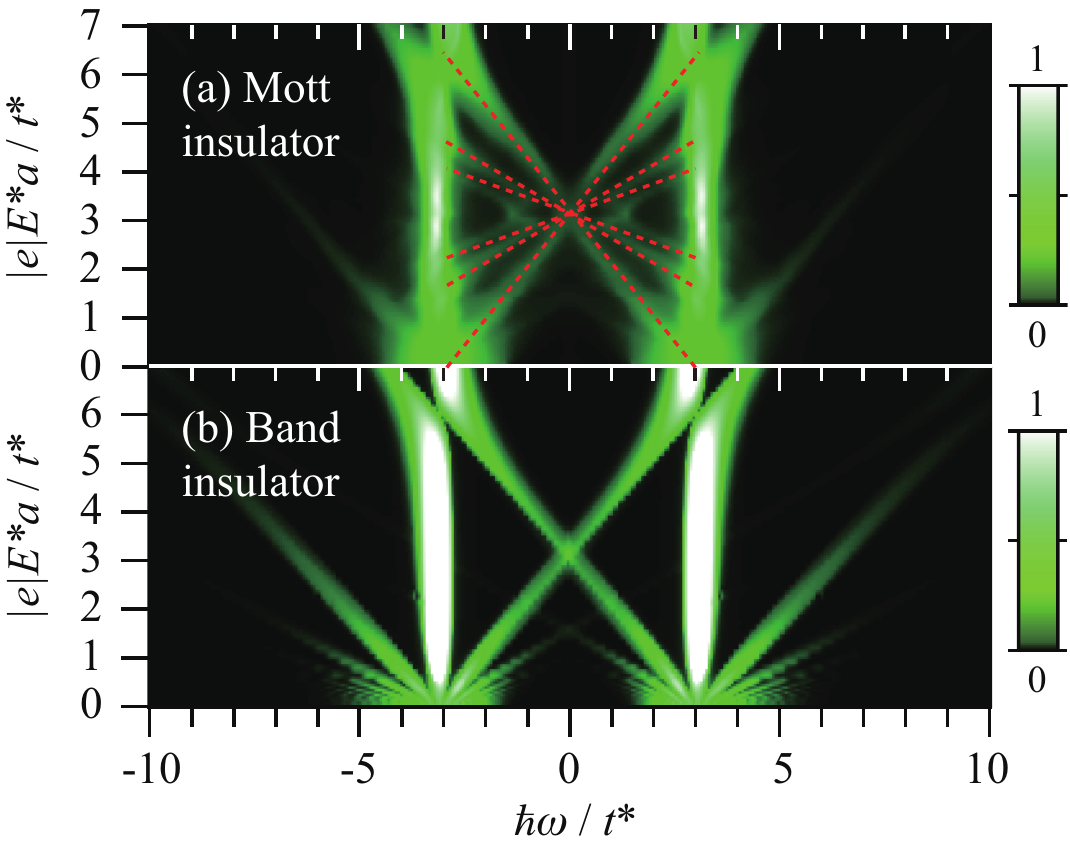} \\
\caption{
Local DOS of the electric-field-driven (a) Mott and (b) noninteracting band insulator.
In sharp contrast to the Mott insulator, where effective WSL branches are shown in red dotted lines, the noninteracting band insulator does not show any signature for the emergent WSL structure.
Parameters for the band insulator are such that (i) the band gap is $6 t^*$ and (ii) the inter-orbital hopping amplitude is the same as the intra-orbital counterpart. 
$V_\textrm{imp} / t^* = 0.2$ and $\Gamma/t^*=0.01$ for both insulators.
}
\label{Fig4_Butterfly}
\end{figure}

What is novel in this work is that, albeit missing the central branch, the reconstructed mid-gap state generates its own emergent WSL branches, whose spacing is roughly proportional to an effective electric field: $E^*_\mathrm{eff} = E^* - E^*_\mathrm{mer}$.
Specifically, the emergent WSL branches are described by $\hbar\omega \simeq n \hbar(\Omega-\Omega_\textrm{mer})$ with $n$ being a non-zero integer and $\hbar\Omega_\textrm{mer}=|e|E^*_\textrm{mer} a$. 
See the comparison between the actual emergent WSL branches and the lines predicted by the above formula, which are plotted in red dotted lines in Fig.~\ref{Fig4_Butterfly}~(a).
It is believed that the central, i.~e., $n=0$, WSL branch vanishes due to the presence of a residual repulsive interaction, which suppresses the local DOS at $\omega=0$ except at $E^* \simeq E^*_\textrm{mer}$.

While specific situations are very different, it is interesting to mention that there is an intriguing parallel between our discovery and the composite fermion theory~\cite{Jain07} for the fractional quantum Hall effect, where strong correlation is renormalized away with exchange of the electric and magnetic field, respectively.
Composite fermions are formed in the lowest Landau level in order to minimize the Coulomb interaction cost between electrons by attaching correlation holes to each electron.
In the lowest Landau level, the holomorphic constraint enforces correlation holes to take the same wave function form as quantized vortices, which, in combination of Fermi statistics, leads to the formation of the bound state between electron and even number of, or most prominently two vortices. 
At half filling, i.~e., at $B=B_{1/2}=2\rho\phi_0$, where $B$ is the strength of the external magnetic field, $\rho$ is the electron density, and $\phi_0$ is the magnetic flux quantum, the total number of vortices captured by composite fermions is exactly equivalent to the entire external magnetic field in units of magnetic flux quantum. 
In this situation, weakly-interacting composite fermions do not experience any effective magnetic field and therefore form an effective Fermi sea state.  
In the vicinity of half filling, composite fermions experience an effective magnetic field, $B_\textrm{eff}$, which is offset from the external magnetic field by $B_{1/2}$: $B_\textrm{eff}=B-B_{1/2}$.

The specific setting of our problem for the electric-field-driven Mott insulator is different from the above. 
However, the interplay between magnetic field and correlation in the composite fermion theory is strongly reminiscent of that between electric field and correlation in our problem. 
That is, in our problem, the local DOS exhibits an effective WSL structure near $E^*=E^*_\textrm{mer}$ as if there is an effective electric field, $E^*_\textrm{eff}$, which is offset from the external electric field by $E^*_\textrm{mer}$: $E^*_\textrm{eff}=E^*-E^*_\textrm{mer}$.
Further analyses are necessary to resolve if there is indeed an underlying reason beyond the superficial similarity.

To confirm that the emergent WSL structure is induced truly by the many-body correlation effect, in Fig.~\ref{Fig4_Butterfly}, we make a comparison between the local DOS of the electric-field-driven Mott insulator and that of the noninteracting {\it band} insulator with two on-site orbitals. 
For convenience, we set the band gap of the noninteracting band insulator to be equal to the Mott gap.  
See Appendix~\ref{appen:band_insulator} for details regarding the computation of the local DOS for the electric-field-driven band insulator.
As one can see, two WSL branches respectively emanating from the upper and lower noninteracting band also meet at $\omega=0$ around $E^*_\textrm{mer}$, but, in sharp contrast to the Mott insulator, simply passes through without generating the emergent WSL structure.
From this comparison, it is clear that the emergent WSL is a consequence of the intricate interplay between strong correlation and large electric field, not a simple interference between sub-bands.

\subsubsection{Population inversion}
\label{sec:inversion}

It is shown in the preceding section that the local DOS exhibits an intriguing pattern of the emergent WSL structure near $E^* = E^*_\textrm{mer} \simeq U/2|e|a$, where two WSL branches respectively emanating from the upper and lower Hubbard band merge at $\omega=0$.  
It is observed in Fig.~\ref{Fig3_Spectrum}~(b) that there is also another intriguing pattern in the local occupation number, which reveals an additional peculiarity of the nonequilibrium steady states arising from the electric-field-driven Mott insulator.

The behavior of the local occupation number undergoes a qualitative change around  $E^*=E^*_\textrm{mer}$.
First, for $E^* \lesssim E^*_\textrm{mer}$, basically all states below $\omega=0$, which include the original WSL branches emanating from the lower Hubbard band and the emergent WSL branches emanating left downward from $E^*=E^*_\textrm{mer}$, are occupied, as accentuated by the corresponding local distribution function in Fig.~\ref{Fig3_Spectrum}~(c).
Note that the local distribution function is described basically by the usual Fermi-Dirac distribution function (with the chemical potential residing at $\omega=0$) except in a narrow region near $\omega=0$, where its value fluctuates around $1/2$.  
In particular, at $E^* = E^*_\textrm{mer}$, the local distribution function becomes more or less flat over a wide range of energy.   
It is interesting to mention that one can define an effective temperature by fitting the local distribution function to the Fermi-Dirac distribution function in the vicinity of $\omega=0$. 
In this definition, the effective temperature becomes infinite at $E^* = E^*_\textrm{mer}$.
It is important to note, however, that the actual temperature, dictated by the thermal bath via the Markovian quantum master equation, is always set to be zero in this work regardless of the parameter regime.

Second, for $E^* \gtrsim E^*_\textrm{mer}$, the local occupation number shows a drastic change in its behavior in the sense that there is now an inversion of the occupation number for the same local DOS, which means that the distribution function gets larger in a higher energy.
This phenomenon is called the {\it population inversion}.
Specifically, the emergent WSL branches emanating right upward from $E^*=E^*_\textrm{mer}$ are more occupied than those emanating left upward.
To appreciate more precisely where the population inversion occurs, see Fig.~\ref{Fig3_Spectrum}~(c), which shows the evolution of the local distribution function as a function of electric field.
As one can see, the local distribution function develops quite a complex, non-monotonic pattern upon entering the regime of $E^* \gtrsim E^*_\textrm{mer}$.
The population inversion is very unusual in condensed matter systems.
As shown by the direct-current analysis in Sec.~\ref{sec:DC_current}, the population inversion eventually causes an instability of the uniform nonequilibrium steady state at sufficiently weak electron-impurity interaction.

\subsection{Direct current}
\label{sec:DC_current}

\subsubsection{Exact formula}
\label{sec:exact_formula}

Now, we move to the direct-current analysis.
Usually, the current is computed in the linear response theory via conductivity, which requires an evaluation of the current-current correlation function with an appropriate approximation for the vertex correction. 
Fortunately, in our problem, we do not have to make any approximations so long as the lesser Green's function is obtained as a function of electric field for general strength. 
The key to this fortunate situation is the fact that the steady-state direct current can be expressed in terms of the lesser Green's function via an exact formula, which is derived in this section.

We begin by writing the current density operator, which is the product of the electron charge and the group velocity operator:
\begin{align}
\mathbf{J}(t) 
= -|e| \sum_\mathbf{k} \mathbf{v}_{\mathbf{k}(t)} 
= -|e| \sum_{\mathbf{k},\sigma} \frac{\partial \epsilon_{\mathbf{k}(t) }}{\partial \hbar \mathbf{k}} c^\dag_{\mathbf{k} \sigma} c_{\mathbf{k} \sigma} ,
\end{align}
where $\partial \epsilon_{{\bf k}(t)} / \partial \hbar{\bf k}$ is the group velocity in the dynamical vector potential gauge with
$\mathbf{k}(t)=\mathbf{k}-|e|{\bf E}t/\hbar$.
Now, since the current should be parallel to the field direction (unless there is a topological term causing transverse components~\cite{Xiao10}), we focus on the normalized current density defined by
$\mathcal{J}^*(t) = \langle \mathbf{J}(t) \rangle \cdot \mathbf{E} / E^*$, which is written as follows:
\begin{align}
\mathcal{J}^*(t)
= & -|e| \sum_{\mathbf{k},\sigma} \sum_{j=1}^d \frac{\partial \epsilon_{\mathbf{k}(t)}}{\partial \hbar k_j} \langle c^\dag_{\mathbf{k} \sigma}(t) c_{\mathbf{k} \sigma}(t) \rangle 
\nonumber \\
= &~ \frac{2i|e|a}{\hbar} \int_0^{2\pi} d\theta_\mathbf{k} \int_0^\infty d\zeta_\mathbf{k} ~\zeta_\mathbf{k}^2 ~\rho_{\mathrm{non}}(\zeta_\mathbf{k}) \sin(\Omega t - \theta_\mathbf{k}) 
\nonumber \\
&\times \hbar G^<(\zeta_\mathbf{k},\theta_\mathbf{k},t_\mathrm{rel} = 0, t_\mathrm{av} = t),
\label{ExactFormula}
\end{align}
where we use that
$\langle c^\dag_{\mathbf{k} \sigma}(t) c_{\mathbf{k} \sigma}(t) \rangle = -i \hbar G^<_{\mathbf{k}\sigma}(t_\mathrm{rel} = 0, t_\mathrm{av} = t)= -i \hbar G^<(\zeta_{\bf k}, \theta_{\bf k}, t_\mathrm{rel} = 0, t_\mathrm{av} = t)$. 
The spin index, $\sigma$, is dropped in the above since we are interested in the paramagnetic phase, where the spin summation simply produces an overall factor of 2. 
Also, it is used in the above that
$\sum_{j=1}^d \frac{\partial \epsilon_{{\bf k}(t)}}{\partial \hbar k_j}=\frac{a}{\hbar}\zeta_{\bf k}\sin{(\Omega t -\theta_{\bf k})}$.
As before, 
$\rho_\mathrm{non}(\zeta_\mathbf{k}) = \exp(-\zeta_\mathbf{k}^2 / {t^*}^2) / \pi {t^*}^2$,
$\zeta_\mathbf{k}=\sqrt{\epsilon_\mathbf{k}^2 +\bar{\epsilon}_\mathbf{k}^2}$, and 
$\tan{\theta_\mathbf{k}}=\bar{\epsilon_\mathbf{k}}/\epsilon_\mathbf{k}$ with 
$\epsilon_\mathbf{k}=-\frac{t^*}{\sqrt{d}}\sum_{j=1}^{d} \cos(k_j a)$ and 
$\bar{\epsilon}_\mathbf{k}=-\frac{t^*}{\sqrt{d}}\sum_{j=1}^{d} \sin(k_j a)$.

Focusing on the steady-state limit, where all aperiodic transient responses disappear, we perform the Floquet-mode expansion of the lesser Green's function:
\begin{align}
& \hbar G^<(\zeta_\mathbf{k},\theta_\mathbf{k},t_\mathrm{rel} = 0, t_\mathrm{av} = t) 
\nonumber \\
&= \sum_{m,n=-N_{\rm c}}^{N_{\rm c}} e^{-i(m-n)(\Omega t - \theta_\mathbf{k})} \int_{-\hbar\Omega/2}^{\hbar\Omega/2} \frac{d\hbar\omega}{2\pi} G_{mn}^<(\zeta_\mathbf{k},\theta_\mathbf{k}=0,\omega) 
\nonumber\\
&= \sum_{s=1}^{2N_{\rm c}} \sum_{n=-N_{\rm c}}^{N_{\rm c}-s} 
\int_{-\hbar\Omega/2}^{\hbar\Omega/2} \frac{d\hbar\omega}{2\pi} \bigg[e^{-is(\Omega t - \theta_\mathbf{k})} G_{n+s,n}^<(\zeta_\mathbf{k},\theta_\mathbf{k}=0,\omega) \nonumber\\
&~~~ - e^{is(\Omega t - \theta_\mathbf{k})} G_{n+s,n}^{<*}(\zeta_\mathbf{k},\theta_\mathbf{k}=0,\omega) \bigg] 
\nonumber\\
&~~~ + \sum_{n=-N_{\rm c}}^{N_{\rm c}} \int_{-\hbar\Omega/2}^{\hbar\Omega/2} 
\frac{d\hbar\omega}{2\pi} G_{nn}^<(\zeta_\mathbf{k},\theta_\mathbf{k}=0,\omega).
\label{LesGreen}
\end{align}
In the above, it is crucial to use the fact that 
$G^<_{mn}(\zeta_{\bf k},\theta_{\bf k},\omega)=e^{i(m-n)\theta_{\bf k}}G^<_{mn}(\zeta_{\bf k},\theta_{\bf k}=0,\omega)$.
Inserting Eq.~\eqref{LesGreen} into \eqref{ExactFormula}, followed by the integration over $\theta_\mathbf{k}$, gives rise to the desired, exact formula for the steady-state current density:
\begin{align}
\mathcal{J}^* 
= &~ \mathcal{J}_0^* \sum_{n=-N_{\rm c}}^{N_{\rm c}-1} \int_{-\hbar\Omega/2}^{\hbar\Omega/2} 
\frac{d\hbar\omega}{2\pi} \int_0^\infty d(\zeta_\mathbf{k} / t^*) 
\nonumber \\
&\times
\zeta^2_\mathbf{k} \rho_{\mathrm{non}}(\zeta_\mathbf{k}) 
\mathrm{Re} G_{n+1,n}^<(\zeta_\mathbf{k},\theta_\mathbf{k}=0,\omega),
\label{ExactCurrent}
\end{align}
where $\mathcal{J}_0^* = 4\pi|e|at^* / \hbar$.
It is interesting to note that the integration over $\theta_{\bf k}$ eliminates all time-dependent components, leaving only the direct current.

Finally, we check if Eq.~\eqref{ExactCurrent} is consistent with the clean limit, where there is no elastic scattering at all, and therefore no direct current should flow. 
To check this, we note that, in the absence of elastic scattering, i.~e., $\Sigma^{r,<}=0$,
the full lesser Green's function, $G^{<}_{mn}(\omega)$, reduces simply to the noninteracting counterpart, $g^{<}_{mn}(\omega)$: 
\begin{align}
&g_{mn}^<(\zeta_\mathbf{k},\theta_\mathbf{k}=0,\omega) 
\nonumber \\
&= -{\cal U}(\zeta_{\bf k},\theta_{\bf k}=0) \Big\{  g^r_{\rm lad}(\omega)  {\cal F}(\omega) 
-{\cal F}(\omega) [g^r_{\rm lad}(\omega)]^{\dagger}  \Big\} 
\nonumber \\
&~~~ \times {\cal U}(\zeta_{\bf k},\theta_{\bf k}=0)
\nonumber \\
&= \sum_p J_{m-p}(\zeta_\mathbf{k}/\hbar\Omega) J_{n-p}(\zeta_\mathbf{k}/\hbar\Omega) f_\mathrm{FD}(\omega+p\Omega) 
\nonumber \\
&~~~ \times \frac{i \Gamma}{[\hbar(\omega+p\Omega)]^2 + (\Gamma/2)^2} .
\label{Clean_limit}
\end{align}
As one can see, Eq.~\eqref{Clean_limit} is pure imaginary and therefore Eq.~\eqref{ExactCurrent} vanishes. 
By contrast, it is interesting to mention that the clean limit is not properly captured by the fermion-bath model, where $g_\mathbf{k}^<$ is no longer pure imaginary even in the complete absence of elastic (i.~e., electron-impurity and electron-electron) scatterings.

\begin{figure}
\centering
\includegraphics[width=0.46\textwidth]
{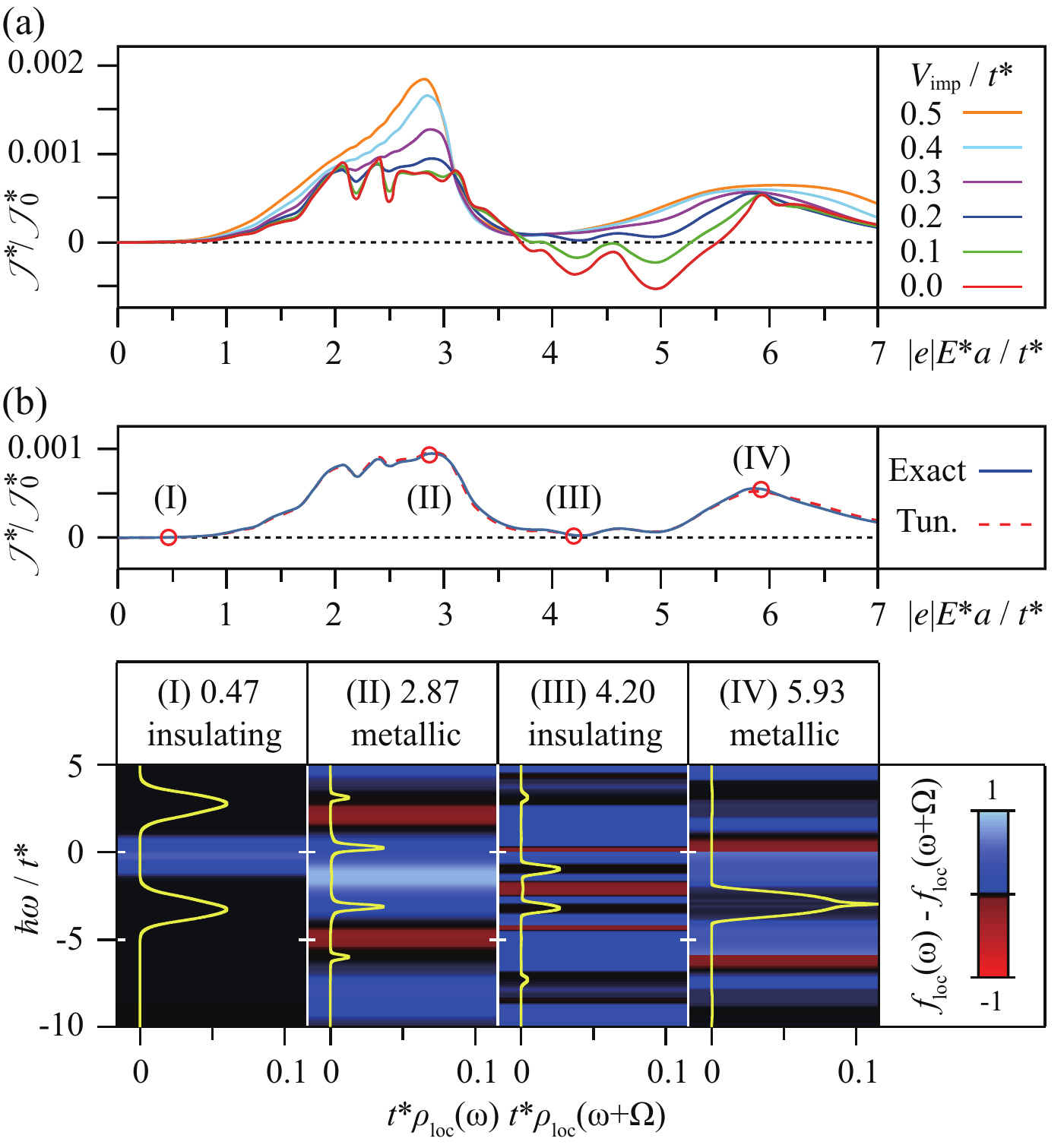} \\
\caption{
(a) Normalized direct-current density, $\mathcal{J}^*$, in units of $\mathcal{J}^*_0=4\pi |e| a t^*/\hbar$ as a function of $E^*$ for the electric-field-driven Mott insulator at $U / t^* = 6$ and various $V_\textrm{imp}/t^*$. As with Fig.~\ref{Fig3_Spectrum} and \ref{Fig4_Butterfly}, $\Gamma/t^*=0.01$.
(b) Comparison between the direct current obtained from the tunneling (red dotted line) and exact (blue solid line) formula at $V_\textrm{imp}/t^*=0.2$.
While the plot only shows the comparison at $V_\textrm{imp}/t^*=0.2$, it is confirmed that the tunneling formula produces excellent agreements with the exact results in the entire studied range of $V_\textrm{imp}/t^*$.
The lower panels provide a graphical explanation for why the direct current is small in the regime of (I) and (III), and large in the regime of (II) and (IV).   
Note that the yellow lines show $\rho_\textrm{loc}(\omega) \rho_\textrm{loc}(\omega+\Omega)$, while the colored backgrounds denote $f_\textrm{loc}(\omega) - f_\textrm{loc}(\omega+\Omega)$ with light blue and red being 1 and -1, respectively. 
}
\label{Fig5_Current}
\end{figure}

\subsubsection{Dielectric breakdown and instability}
\label{sec:dielectric_breakdown}

Figure~\ref{Fig5_Current}~(a) plots the normalized direct-current density, $\mathcal{J}^*$, computed according to Eq.~\eqref{ExactCurrent} as a function of $E^*$ for the Mott insulator at $U / t^* = 6$ and various $V_\textrm{imp}/t^*$.
The behavior of $\mathcal{J}^*$ is overall similar for various $V_\textrm{imp}/t^*$ so long as $V_\textrm{imp}/t^*$ is sufficiently large, say, $V_\textrm{imp}/t^* \gtrsim 0.2$;
(i) initially, the Mott insulator remains insulating at small electric fields, 
(ii) then, $\mathcal{J}^*$ begins to increase as a function of $E^*$ until it hits the first maximum around $E^*_\textrm{mer}$,  
(iii) then, $\mathcal{J}^*$ decreases sufficiently such that the Mott insulator returns to an insulating phase, and 
(iv) finally, $\mathcal{J}^*$ increases again until it reaches the second maximum around $2 E^*_\textrm{mer}$, and slowly decays roughly as $1/E^*$.
In summary, for sufficiently large electron-impurity scattering, the electric-field-driven Mott insulator undergoes a sequence of two dielectric breakdowns mediated by a reentrant insulating phase in the middle.

For weak electron-impurity scattering, say, $V_\textrm{imp}/t^* \lesssim 0.2$, the situation becomes different, especially more for the reentrant insulating phase. 
While maintaining the overall structure of two dielectric breakdowns, $\mathcal{J}^*$ shows many additional short-period oscillations as a function of $E^*$. 
In the reentrant insulating regime, these short-period oscillations make $\mathcal{J}^*$ fluctuate so much that $\mathcal{J}^*$ is not necessarily bounded as a positive value. 
The apparently unphysical negative $\mathcal{J}^*$ means that the current flows against the applied electric field, which is believed to cause an instability of the spatially-uniform steady state toward inhomogeneous current density states. Such behavior was argued by Andreev and collaborators~\cite{Andreev03} in the context of the radiation-induced zero-resistance state observed in the quantum Hall system~\cite{Zudov01,Mani02,Zudov03,Shi03,Durst03,Andreev03,Park04,Durst04}. 

It is important to note that the instability occurs in the reentrant insulating regime, where the population inversion is observed. 
In the following section, we show that the coincidence between the regime of instability and population inversion is not by chance.
Specifically, we derive the tunneling formula to show that the negative current can be obtained if the local DOS and distribution function satisfy delicate conditions. 
A necessary condition is the population inversion.

\subsubsection{Tunneling formula}
\label{sec:tunneling_formula}

It has been shown in the preceding section that the direct current can be computed exactly according to Eq.~\eqref{ExactCurrent}, if the lesser Green's function is given as a function of electric field for general strength.   
While this computation is technically sufficient, it is desirable that there is a physical connection between the local spectrum computed in Sec.~\ref{sec:local_spectrum} and the direct current.

Inspired by the observation that our system may be viewed as a tunneling junction array of multiple quantum dots with a voltage bias applied between quantum dots, we consider the following tunneling formula:
\begin{align}
\mathcal{J}^*_\mathrm{tun} =& \mathcal{J}^*_{\mathrm{tun},0} \int^{\infty}_{-\infty} d(\hbar\omega/t^*) \; t^* \rho_\mathrm{loc}(\omega) t^* \rho_\mathrm{loc}(\omega+\Omega)
\nonumber \\
&\times 
\left[ f_\mathrm{loc}(\omega) - f_\mathrm{loc}(\omega+\Omega) \right] ,
\label{eq:tunneling}
\end{align}
where $\mathcal{J}^*_{\mathrm{tun},0}$ is an overall constant.
Postponing the justification of the tunneling formula for the time being, let us first focus on Fig.~\ref{Fig5_Current}~(b), which shows the comparison between the normalized direct-current density obtained from the exact formula, $\mathcal{J}^*$, and that from the tunneling formula, $\mathcal{J}^*_\mathrm{tun}$, at $V_\textrm{imp}/t^*=0.2$.
As one can see, the agreement is excellent, strongly suggesting an existence of the rigorous derivation for the tunneling formula.

Indeed, it is shown in Appendix~\ref{appen:tunneling_formula} that the above tunneling formula can be derived rigorously by expanding $[G^<_{{\bf k}}(\omega)]_{mn}$ in Eq.~\eqref{ExactCurrent} up to first order of $\zeta_{\bf k}$, which, formally, amounts to taking the limit of weak tunneling, i.~e., $t^* \ll \hbar \Omega$, $U$, $V_\textrm{imp}$, for a given fixed self-energy. 
It is important to note that the tunneling formula can become valid beyond the strict weak-tunneling limit since $\rho_\textrm{loc}$ as well as $f_\textrm{loc}$ by themselves contain entire effects of tunneling via the self-energy, which is obtained as a converged solution of the DMFT self-consistency loop.  
A useful bonus of the rigorous derivation is that $\mathcal{J}^*_{\mathrm{tun},0}$ is now precisely obtained as $\mathcal{J}^*_0/4$.
See Appendix~\ref{appen:tunneling_formula} for details.

The remarkable success of the tunneling formula provides a valuable physical insight toward the origin of the dielectric breakdowns.
One can see from the tunneling formula that a necessary condition should be satisfied in order for the direct current to become finite, i.~e., for the dielectric breakdown to occur.
The necessary condition is that $\rho_\textrm{loc}(\omega)$ should have a sizable overlap with $\rho_\textrm{loc}(\omega+\Omega)$ within the nonzero window of $f_\textrm{loc}(\omega)-f_\textrm{loc}(\omega+\Omega)$ (which is not necessarily positive anymore).

It is shown in the lower panels of Fig.~\ref{Fig5_Current}~(b) that
the first dielectric breakdown occurs at $E^* = E^*_\textrm{mer} \simeq U/2|e|a$ since the reconstructed mid-gap state at $\omega=0$ has a finite local DOS overlap with the Hubbard band centered at $\omega \simeq \Omega$ within the positive window of $f_\textrm{loc}(\omega)-f_\textrm{loc}(\omega+\Omega)$.
Meanwhile, the second dielectric breakdown occurs at $E^* = 2 E^*_\textrm{mer}$ since the lower Hubbard band has a finite local DOS overlap with the upper counterpart. 
In summary, the first dielectric breakdown is induced by a coherent reconstruction of the mid-gap state within the Mott gap, while the second is induced by an incoherent tunneling between the lower and upper Hubbard band. 

Finally, by using the tunneling formula, one can explain how the population inversion can generate the negative direct current. 
To this end, let us note that the population inversion means that $f_\textrm{loc}(\omega)-f_\textrm{loc}(\omega') < 0$ for $\omega < \omega'$.
The negative direct current can be generated if $f_\textrm{loc}(\omega)-f_\textrm{loc}(\omega+\Omega) < 0$ in the region where $\rho_\textrm{loc}(\omega)\rho_\textrm{loc}(\omega+\Omega)$ is sizable.

\section{Conclusion}
\label{sec:conclusion}

In this work, we perform systematic analyses for the evolution of the nonequilibrium steady states arising from the electric-field-driven Mott insulator. 
As the main theoretical framework, we use the Keldysh-Floquet dynamical mean field theory (DMFT), which can, in principle, treat strong correlation in general nonequilibrium situations. 
The Keldysh-Floquet DMFT is implemented, based on the standard Keldysh-contour expansion, via which the fully-interacting (retarded as well as lesser) Green's functions are connected with the noninteracting counterparts in the infinite past, at which the interaction between electrons is switched on.
Our approach is distinguished from previous ones in a very important aspect that the electric field is assumed to be always on, or turned on even prior to the switching of the interaction between electrons.  
In this situation, we first solve the noninteracting Hamiltonian in the presence of a static electric field to obtain the exact eigenstates, i.~e., the Wannier-Stark ladder (WSL) eigenstates.

A crucial breakthrough in this work is to show that the WSL eigenstates are individually thermalized via the standard canonical ensemble. 
We prove this by solving the Markovian quantum master equation, which is derived for the Caldeira-Leggett model with the coupling being existent between the system and a bosonic thermal bath. 
With the so-obtained noninteracting Green's functions plugged into the Keldysh-Floquet DMFT self-consistency loop as an input, we then obtain the fully-interacting Green's functions by solving the Keldysh-Dyson equation, which is embedded in the Keldysh-Floquet DMFT self-consistency loop.
The self-energy is determined via the iterated perturbation theory (IPT), which is the impurity solver of this work.

From the converged solution of the Keldysh-Floquet DMFT self-consistency loop, we compute the local spectral quantities such as the local density of states (DOS) and distribution function. 
The direct current is computed rigorously with help of an exact formula derived in Sec.~\ref{sec:exact_formula}.  
One of the most important insights obtained in this work is that the direct-current response of the electric-field-driven Mott insulator is closely related with its local spectral properties captured by the local DOS and distribution function. 
Their close relationship is demonstrated in Sec.~\ref{sec:tunneling_formula} through the fact that the direct current computed via the exact formula is reproduced very accurately by the tunneling formula, which is derived rigorously in the weak-tunneling limit.

The success of the tunneling formula enables us to draw the conclusion that the dielectric breakdown occurs in two separate processes with the first via a coherent reconstruction of the mid-gap state within the Mott gap and the second via an incoherent tunneling through the biased Hubbard bands. 
The reconstructed mid-gap state generates its own emergent WSL structure with a reduced effective electric field, as if strong correlation is renormalized away with exchange of the electric field.    
The two metallic phases are mediated by a reentrant insulating phase, which is characterized by the population inversion, causing instability toward inhomogeneous current density states at weak electron-impurity scattering.

It is important to note that the local spectral properties of the electric-field-driven Mott insulator are characterized by two remarkable peculiarities: (i) the appearance of the emergent WSL structure and (ii) the population inversion in the reentrant insulating regime.   
For future work, it would be interesting to investigate various physical implications of these two spectral properties.
First, it is mentioned in Sec.~\ref{sec:emergentWSL} that there is a superficial similarity between the composite fermion theory for the fractional quantum Hall effect and our discovery in this work. 
It would be worthwhile to investigate the physical origin of the emergent WSL structure to see if there is an underlying physical reason beyond the superficial similarity.
Second, it is mentioned in Sec.~\ref{sec:inversion} that the population inversion is related to the negative direct current, which is predicted to occur in the reentrant insulating regime at weak electron-impurity scattering.    
The negative direct current is thermodynamically unstable against the appearance of inhomogeneous current density states.
It would be interesting to investigate exactly how the inhomogeneous current density states can arise. To achieve this goal, a theoretical framework beyond the DMFT might be required.

Finally, the nature of the nonequilibrium steady states arising from the electric-field-driven Mott insulator depends crucially on the thermalization of the noninteracting system, which is determined in this work via the Markovian quantum master equation.
It would be interesting to investigate if other thermalization schemes are possible. Also, in this work, we have focused only on the diagonal components of the Lindblad equation, since only the diagonal components survive in the steady-state limit.
It would be interesting to study the dynamics of the off-diagonal components as well, which can tell us about the decoherence of the system.

\begin{acknowledgments}
The authors are grateful to Jaeyoon Cho for his valuable comments on our initial model for the energy dissipation, to Mahn-Soo Choi and Myung-Joong Hwang for their helpful comments regarding the derivation of the Markovian quantum master equation, and to Jong Eun Han for his careful explanation of the fermion-bath model. 
In particular, the authors are indebted to Jainendra Jain for the discussion regarding the parallel between the formation of the emergent Wannier-Stark ladder and the composite fermion theory.
Also, the authors thank KIAS Center for Advanced Computation (CAC) for providing computing resources.
This research was supported in part by the National Research Foundation of Korea (NRF) funded by the Korea government (MEST) under Quantum Metamaterials Research Center, Grant No. 2008-0062238 (K.P.).
\end{acknowledgments}

\appendix

\section{Computation of the noninteracting Joint DOS}
\label{appen:jointDOS}

In this section, we present details for the computation of the noninteracting joint DOS, $\rho_{\rm non}(\zeta)$.
We begin from the definition of $\rho_{\rm non}(\zeta)$:
\begin{align}
&\rho_{\rm non} (\epsilon,\bar{\epsilon}) 
\nonumber \\
&= \int \frac{d^d (ka)}{(2\pi)^d} \delta(\epsilon-\epsilon_{\bf k})\delta(\bar{\epsilon}-\bar{\epsilon}_{\bf k})
\nonumber \\
&= \int \frac{ds}{2\pi} \int \frac{d\bar{s}}{2\pi} \int \frac{d^d (ka)}{(2\pi)^d} e^{is(\epsilon-\epsilon_{\bf k})} e^{i\bar{s}(\bar{\epsilon}-\bar{\epsilon}_{\bf k})}
\nonumber \\
&= \int \frac{ds}{2\pi} e^{is\epsilon} \int \frac{d\bar{s}}{2\pi}  e^{i\bar{s}\bar{\epsilon}} \int \frac{d^d (ka)}{(2\pi)^d}  
e^{i\frac{t^*}{\sqrt{d}} \sum^{d}_{j=1}( s \cos{k_j a} +\bar{s}\sin{k_j a} ) }
\nonumber \\
&= \int \frac{ds}{2\pi} e^{is\epsilon} \int \frac{d\bar{s}}{2\pi}  e^{i\bar{s}\bar{\epsilon}} 
\left[
\int^{\pi}_{-\pi} \frac{d (ka)}{2\pi}  
e^{i\frac{t^*}{\sqrt{d}} \sqrt{s^2+\bar{s}^2} \cos{(k_j a -\theta_s)} }
\right]^d 
\nonumber \\
&= \int \frac{ds}{2\pi} e^{is\epsilon} \int \frac{d\bar{s}}{2\pi}  e^{i\bar{s}\bar{\epsilon}} 
\left[ J_0 \left(\frac{t^*}{\sqrt{d}}\sqrt{s^2+\bar{s}^2}\right) \right]^d ,
\label{jointDOS_d}
\end{align}
where $\tan{\theta_s}=\bar{s}/s$ and $J_0(x)$ is the zeroth Bessel function of the first kind. 
Note that, in the second line, we use the fact that $\delta(x-a) = \int \frac{ds}{2\pi} e^{is(x-a)}$.

In the limit of infinite dimensions, i.~e., $d \rightarrow \infty$, the argument of the Bessel function goes to zero, in which case Eq.~\eqref{jointDOS_d} is simplified as follows:
\begin{align}
&\rho_{\rm non} (\epsilon,\bar{\epsilon}) 
\nonumber \\
&= \int \frac{ds}{2\pi} e^{is\epsilon} \int \frac{d\bar{s}}{2\pi}  e^{i\bar{s}\bar{\epsilon}} 
\left[ 1 +\frac{t^{*2}}{4d} (s^2+\bar{s}^2) +{\cal O}\left( \frac{1}{d^2} \right) \right]^d , 
\nonumber \\
&= \int \frac{ds}{2\pi} e^{is\epsilon} \int \frac{d\bar{s}}{2\pi}  e^{i\bar{s}\bar{\epsilon}} 
\exp{\left( t^{*2}(s^2+\bar{s}^2)/4  +{\cal O}\left( \frac{1}{d} \right)  \right)}
\nonumber \\
&\approx \frac{1}{\pi t^{*2}}
\exp{\left(  
-\frac{\epsilon^2+\bar{\epsilon}^2}{t^{*2}}
\right)}
\nonumber \\
&= \frac{1}{\pi t^{*2}}
\exp{\left(  
-\frac{\zeta^2}{t^{*2}}
\right)},
\label{jointDOS}
\end{align}
where we use the fact that $J_0(x) \approx 1 -(x/2)^2$ for $x \ll 1$.

\section{Energy conservation in the vector potential gauge}
\label{appen:energy_conservation}

In this section, we show that, with proper definition, the total energy is still conserved in the dynamical vector potential gauge.
To this end, we begin by writing the noninteracting Hamiltonian in the dynamical vector potential gauge, $\phi = 0$ and $\mathbf{A} = -c\mathbf{E}t$:
\begin{align}
H(t) = - \sum_{\langle ij \rangle, \sigma} t_{ij} \left[ e^{ i\varphi_{ij}(t)} c_{i\sigma}^\dag c_{j\sigma} + \textrm{H. c.} \right],
\end{align}
where $c_{i\sigma}^\dag$ and $c_{i\sigma}$ $(\sigma = \uparrow, \downarrow)$ are the electron creation and annihilation operator at the $i$-th lattice site, respectively, and $t_{ij}$ is the hopping amplitude. 
The Peierls phase factor is defined via  
$\varphi_{ij}(t) = \frac{|e|}{\hbar c} \int_{\mathbf{r}_i}^{\mathbf{r}_j}\mathbf{A}(t) \cdot d\mathbf{r} =-\frac{|e|}{\hbar} t \mathbf{E}\cdot(\mathbf{r}_j-\mathbf{r}_i)$, 
where $e(<0)$ is the charge of electron. 
In the momentum space, the Hamiltonian is simplified as follows:
\begin{align}
H(t) = \sum_{\mathbf{k}\sigma} \epsilon_{\mathbf{k}(t)} c_{\mathbf{k}\sigma}^\dag c_{\mathbf{k}\sigma} ,
\end{align}
where $\epsilon_\mathbf{k} = -\sum_{\langle i, 0 \rangle} t_{i0} e^{-i\mathbf{k} \cdot (\mathbf{r}_i  -\mathbf{r}_0)}$ with $\mathbf{r}_0$ being a reference position vector and
$\mathbf{k}(t) =\mathbf{k} + \frac{|e|}{\hbar c}\mathbf{A}(t) = \mathbf{k} - \frac{|e|}{\hbar}\mathbf{E}t$.
Note that the effects of the static electric field are incorporated through the Peierls substitution: 
$\mathbf{k} \rightarrow \mathbf{k}(t)$.

\begin{figure*}
\centering
\includegraphics[width=0.70\textwidth]
{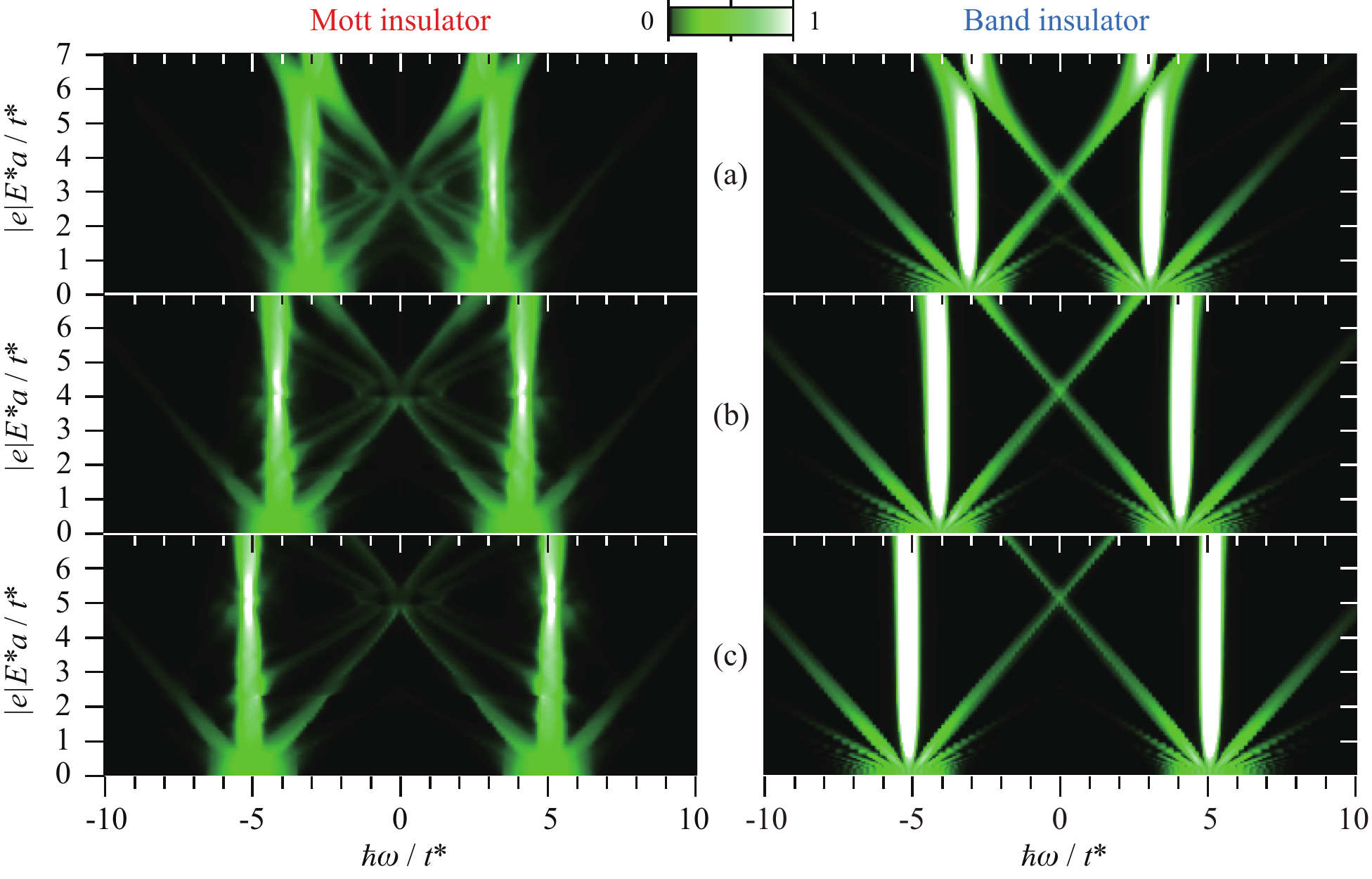} \\
\caption{
Comparison between the local DOS for the Mott and band insulator with the gap size being (a) $6 t^*$, (b) $8 t^*$, and (c) $10 t^*$.
It is shown that, as $U$ increases, the local DOS patterns for both insulators become more similar to each other with the emergent WSL structure getting fainter, indicating that the IPT is consistent with strong-$U$ limit.
Here, we set $V_{\rm imp}/t^*=0.2$ and $\Gamma/t^*=0.01$ for both insulators.
The inter-orbital hopping amplitude is chosen to be the same as the intra-orbital counterpart.
}
\label{Fig6_Appen}
\end{figure*}

To understand the physical meaning of the Hamiltonian and its time evolution, we consider the Heisenberg equation for the Hamiltonian:
\begin{align} 
\frac{d}{dt} H(t) = \frac{1}{i\hbar} [H(t), H(t)]  + \frac{\partial}{\partial t} H(t) 
= - |e| \mathbf{E}\cdot   \sum_\mathbf{k} \mathbf{v}_{\mathbf{k}(t)} ,
\label{eq:dH_dt}
\end{align}
where $\mathbf{v}_{\mathbf{k}(t)}=\frac{\partial \epsilon_{\mathbf{k}(t)}}{\partial \hbar \mathbf{k}} c^{\dagger}_\mathbf{k} c_\mathbf{k}$ is the group velocity operator. 
Since the group velocity can be taken as a physical velocity, it is natural to expect that the group velocity operator should be related to the time derivative of the center-of-mass position operator, ${\bf R}=\sum_{i,\sigma} \mathbf{r}_i c^\dagger_{i\sigma}c_{i\sigma}$.
To check this, let us write the Heisenberg equation for $\mathbf{R}$:
\begin{align} 
\frac{d}{dt} \mathbf{R} &= \frac{1}{i\hbar} [\mathbf{R}, H(t)] 
\nonumber \\
&= \frac{1}{i\hbar} \sum_{i} \sum_{m,n} \mathbf{r}_i 
\tilde{t}_{mn} [c^\dagger_i c_i, c^\dagger_m c_n ] 
\nonumber \\
&=\frac{1}{i\hbar} \sum_{m,n} \tilde{t}_{mn} \left( \mathbf{r}_m -\mathbf{r}_n \right) c^\dagger_m c_n,
\label{eq:dR_dt}
\end{align}
where $\tilde{t}_{mn}=-t_{mn} e^{i\varphi_{mn}(t)}$ with $\tilde{t}_{mn}=\tilde{t}^*_{nm}$.
In the above, the nearest-neighbor restriction for summation indices, $m$ and $n$, is lifted, reflecting the fact that Eq.~(\ref{eq:dR_dt}) is valid for arbitrary $\tilde{t}_{mn}$.  
The last expression in Eq.~(\ref{eq:dR_dt}) is obtained by using the commutation relationship: 
\begin{align}
[c^\dagger_i c_i, c^\dagger_m c_n] = (\delta_{im}-\delta_{in}) c^\dagger_m c_n .
\end{align}
Performing Fourier transformation of Eq.~(\ref{eq:dR_dt}) gives rise to the desired relationship between $d \mathbf{R} / d t$ and $\mathbf{v}_\mathbf{k}$:
\begin{align}
\frac{d}{dt} \mathbf{R} = \sum_\mathbf{k} \mathbf{v}_{\mathbf{k}(t)},
\end{align} 
which in turn simplifies Eq.~(\ref{eq:dH_dt}) as follows:
\begin{align} 
\frac{d}{dt} H(t) = - |e| \mathbf{E}\cdot \frac{d}{dt} \mathbf{R} ,
\end{align}
or equivalently
\begin{align} 
H(t) - e \mathbf{E}\cdot \mathbf{R} = \textrm{Const.},
\label{eq:CoM}
\end{align}
where $e<0$ according to our convention.
Equation~(\ref{eq:CoM}) describes a constant of motion in the dynamical vector potential gauge, which is identical to the total energy in the static scalar potential gauge, i.~e., the sum of the kinetic and electric potential energy.

Therefore, we conclude that the Hamiltonian in the dynamical vector potential gauge is simply the kinetic energy part of the total energy that is conserved according to Eq.~(\ref{eq:CoM}). 
Finally, it is important to note that the total energy of the electron system as a whole is conserved even in the presence of electron-electron interaction since the interaction only transfers the energy from one electron to another, leaving the total sum intact.

\section{Further justification of the iterated perturbation theory}
\label{appen:IPT}

To provide a further justification of the IPT in the strong-$U$ limit, it is necessary to know how the fully-interacting retarded Green's function in Eq.~\eqref{RetGreenLargeU} in Sec.~\ref{sec:IPT} is modified in the presence of a finite, but small hopping amplitude.
To this end, we consider a two-orbital band insulator, whose band gap is equal to $U$.
The rationale behind this consideration is that, since the Mott insulator can be regarded as being composed of two, entirely independent, noninteracting bands in the large-$U$ limit, the next natural step is to investigate the effects of hopping for such effective band insulator. 
With computational details given in Sec.~\ref{appen:band_insulator}, here, we only show the comparison between the resulting local DOS for the Mott and the noninteracting band insulator at several different values of $U$ in Fig.~\ref{Fig6_Appen}.

It is shown in Fig.~\ref{Fig6_Appen} that, as $U$ increases, the local DOS patterns for both insulators become more and more similar to each other while the emergent WSL structure gets fainter in the Mott insulator. 
This indicates that our calculation based on the IPT impurity solver is consistent with what is obtained via essentially an analytical method for the two-orbital band insulator. 
It is, therefore, concluded that the IPT is consistent with the strong-$U$ limit beyond the simple limit of zero hopping amplitude.

\section{Local DOS for the electric-field-driven noninteracting band insulator}
\label{appen:band_insulator}

In this section, we describe how to compute the local DOS for the electric-field-driven, noninteracting band insulator with two on-site orbitals.
Let us begin by writing the Hamiltonian:
\begin{align}
H_\mathrm{BI} &= H_\mathrm{BI, 0} + H_\mathrm{imp} ,
\\
H_\mathrm{BI, 0} &= 
-\sum_{\langle ij \rangle,\sigma,\alpha,\beta} 
t_{ij}^{\alpha\beta} \left[ e^{i\varphi_{ij}(t)} c_{i\sigma\alpha}^\dag c_{j\sigma\beta} + \textrm{H. c.} \right] 
\nonumber \\
&~~~ +\sum_{i,\alpha,\sigma} \epsilon_\alpha  c_{i\sigma\alpha}^\dag c_{i\sigma\alpha} ,
\label{eq:H_BI0}
\\
H_\mathrm{imp} &= V \sum_{i,\sigma,\alpha} n_{i\sigma\alpha} n_{i,\mathrm{imp}} ,
\label{eq:H_BI}
\end{align}
where $c^\dagger_{i\sigma\alpha}$ and $c_{i\sigma\alpha}$  are the electron creation and annihilation operator for spin $\sigma$ $(=\uparrow, \downarrow)$ and orbital $\alpha$  $(=A,B)$ at the $i$-th lattice site, respectively.
For simplicity, the hopping amplitude, $t_{ij}^{\alpha\beta}$, is chosen such that  $t_{ij}^{AA}=t_{ij}^{BB} = t^*/2\sqrt{d}$ and $t_{ij}^{BA}=t_{ij}^{AB} = \gamma t^*/2\sqrt{d}$, with $\gamma$ being the ratio between the inter- and intra-orbital hopping amplitude and $d$ being the spatial dimensions of the system. 
The band gap is given by $\epsilon_{B} -\epsilon_{A} =\Delta_\textrm{band}$. 
As with the main text, $V$ is the on-site electron-impurity interaction strength and $n_{i,\mathrm{imp}}$ is the impurity number operator at the $i$-th lattice site.

Let us first examine the impurity-free part of the Hamiltonian in Eq.~(\ref{eq:H_BI0}), which is written in the momentum space as follows:
\begin{align}
&H_\mathrm{BI, 0} 
\nonumber \\
&=\sum_{\mathbf{k},\sigma}
\left(
\begin{array}{cc}
c^\dagger_{\mathbf{k} \sigma A}  & c^\dagger_{\mathbf{k} \sigma B}
\end{array}
\right) 
\left(
\begin{array}{cc}
\epsilon_{\mathbf{k}(t)}  +\epsilon_A &  \gamma \epsilon_{\mathbf{k}(t)} \\
\gamma \epsilon_{\mathbf{k}(t)}  &  \epsilon_{\mathbf{k}(t)} +\epsilon_B
\end{array}
\right) 
\left(
\begin{array}{c}
c_{\mathbf{k} \sigma A}  \\ 
c_{\mathbf{k} \sigma B}
\end{array}
\right) ,
\label{eq:H_BI_k}
\end{align}
where $\epsilon_\mathbf{k}=-\frac{t^*}{\sqrt{d}}\sum_{i=1}^{d} \cos{k_i a}$ and $\mathbf{k}(t)=\mathbf{k}-\frac{e}{\hbar c} \mathbf{A}(t)=\mathbf{k}-\frac{|e|}{\hbar} \mathbf{E} t$.
From Eq.~(\ref{eq:H_BI_k}), the equation of motion for $c_{\mathbf{k}\sigma\alpha}$ is obtained as follows:
\begin{align}
\frac{d}{dt} c_{\mathbf{k} \sigma \alpha}= -\frac{i}{\hbar} (\epsilon_\alpha +\epsilon_{\mathbf{k}(t)}) c_{\mathbf{k} \sigma \alpha} -\frac{i}{\hbar} \gamma \epsilon_{\mathbf{k}(t)} c_{\mathbf{k} \sigma \bar{\alpha}} ,
\label{dc_dt}
\end{align}
where we use the notation that $\bar{\alpha}$=$A$ if $\alpha$=$B$ and vice versa.
In turn, Eq.~\eqref{dc_dt} gives rise to the equations of motion for the noninteracting retarded Green's function, $g^r_{\mathbf{k}\sigma,\alpha\beta}(t,t')$:
\begin{align}
\frac{\partial}{\partial t} g^r_{\mathbf{k},\alpha\alpha}(t,t')  
&=
-\frac{i}{\hbar} \delta(t-t') -\frac{i}{\hbar} \left< \left\{   \frac{d}{dt} c_{\mathbf{k}\sigma\alpha}(t) , c^\dagger_{\mathbf{k}\sigma\alpha}(t') \right\} \right>
\nonumber \\
&= -\frac{i}{\hbar} \delta(t-t') -\frac{i}{\hbar} (\epsilon_{\mathbf{k}(t)} +\epsilon_\alpha) g^r_{\mathbf{k},\alpha\alpha}(t,t')
\nonumber \\
&~~~ -\frac{i}{\hbar} \gamma\epsilon_{\mathbf{k}(t)} g^r_{\mathbf{k},\bar{\alpha}\alpha}(t,t') ,
\label{eq:dg_dt1}
\end{align}
and
\begin{align}
\frac{\partial}{\partial t} g^r_{\mathbf{k},\bar{\alpha}\alpha}(t,t')  
&= -\frac{i}{\hbar} \left< \left\{   \frac{d}{dt} c_{\mathbf{k}\sigma\bar{\alpha}}(t) , c^\dagger_{\mathbf{k}\sigma\alpha}(t') \right\} \right>
\nonumber \\
&= -\frac{i}{\hbar} (\epsilon_{\mathbf{k}(t)} +\epsilon_\alpha) g^r_{\mathbf{k},\bar{\alpha}\alpha}(t,t')
\nonumber \\
&~~~ -\frac{i}{\hbar} \gamma\epsilon_{\mathbf{k}(t)} g^r_{\mathbf{k},\alpha\alpha}(t,t') ,
\label{eq:dg_dt2}
\end{align}
which are derived from the definition:
\begin{align}
g^r_{\mathbf{k},\alpha\beta}(t,t') = -\frac{i}{\hbar} \theta(t-t') \langle \{  c_{\mathbf{k} \sigma \alpha}(t), c^\dagger_{\mathbf{k} \sigma \beta} (t') \} \rangle .
\end{align}
For simplicity, the spin index, $\sigma$, is dropped in the above since the Green's function does not depend on the spin in the paramagnetic phase.
Rewriting Eqs.~(\ref{eq:dg_dt1}) and (\ref{eq:dg_dt2}) in the Floquet representation gives rise to the following matrix equation for the inverse of $g^r_{\mathbf{k},\alpha\alpha}(\omega)$:
\begin{align}
\big[ g^r_{\mathbf{k},\alpha\alpha}(\omega) \big]^{-1}_{mn} &=
[\hbar(\omega + n\Omega) -\epsilon_\alpha + i\eta] \delta_{mn} - (\epsilon_\mathbf{k})_{mn} 
\nonumber \\
&-\gamma^2 \sum_{p,q} (\epsilon_\mathbf{k})_{mp} 
\big[ g^{r(0)}_{\mathbf{k},\bar{\alpha}\bar{\alpha}}(\omega) \big]_{pq} (\epsilon_\mathbf{k})_{qn}, 
\label{eq:gk_inverse}
\end{align}
where
\begin{align}
\big[ g^{r(0)}_{\mathbf{k},\bar{\alpha}\bar{\alpha}}(\omega) \big]^{-1}_{mn} =
\left[ \hbar(\omega + n\Omega) -\epsilon_{\bar{\alpha}} + i\eta \right] \delta_{mn} 
-(\epsilon_\mathbf{k})_{mn},
\end{align}
where $(\epsilon_\mathbf{k})_{mn} = \int_{-\pi}^{\pi} \frac{dx}{2\pi} e^{i (m-n) x} \epsilon_{\mathbf{k}(x/\Omega)}$.
In the presence of an inelastic scattering, $\eta$ is taken to be $\Gamma/2$ with $\Gamma$ denoting the full broadening width at half maximum.

As with the Mott insulator considered in the main text, the effects of the electron-impurity interaction are addressed within the self-consistent Born approximation (SCBA):
\begin{align}
\Sigma^{r,<}_{\mathrm{imp}, \alpha\alpha}(\omega)=n_\textrm{imp} V^2 G^{r,<}_{\alpha\alpha}(\omega) ,
\label{eq:Sigma_imp}
\end{align}
where $n_\textrm{imp}$ is the average impurity number per site and
$G^{r,<}_{\alpha\alpha}(\omega)$ $[=\sum_\mathbf{k} G^{r,<}_{\mathbf{k}, \alpha\alpha}(\omega)]$ are the local Green's functions, which are self-consistently determined from $g^{r,<}_{\alpha\alpha}(\omega)$ via the Keldysh-Dyson equation.
It is important to note that, within the SCBA, the retarded part of the self-energy is completely independent of the lesser counterpart.

Finally, the local DOS is computed from the imaginary part of the retarded local Green's function with contributions from both orbitals:
\begin{align}
\rho_\textrm{loc}(\omega+n\Omega)=-\frac{1}{\pi}\textrm{Im} \sum_{\alpha=A, B} G^r_{\alpha\alpha, {nn}}(\omega).
\end{align}

\section{Derivation of the tunneling formula}
\label{appen:tunneling_formula}

In this section, we derive the tunneling formula from the exact current formula in Eq.~\eqref{ExactCurrent} by expanding $G^<_{n+1,n}(\zeta_{\bf k},\theta_{\bf k}=0,\omega)$ up to first order of $\zeta_{\bf k}$.
Since $\zeta_\mathbf{k} \propto t^*$, this expansion formally corresponds to taking the limit of weak tunneling, i.~e. $t^* \ll \hbar\Omega$, $U$, $V_\mathrm{imp}$, for a given self-energy that is fixed as a converged solution of the DMFT self-consistency loop.

To perform this expansion, it is convenient to remind ourselves that the diagonal elements of the retarded Green's function have the structure of a continued fraction, as shown in Eq.~\eqref{InvRetGreen_diag}. 
The off-diagonal elements can be determined from the diagonal counterparts via Eq.~\eqref{InvRetGreen_off}.
Then, according to Eqs.~\eqref{InvRetGreen_diag} and \eqref{InvRetGreen_off}, the retarded Green's function can be expanded up to first order of $\zeta_{\bf k}$ as follows:
\begin{align}
G_{nn}^r(\zeta_\mathbf{k},\theta_{\bf k}=0,\omega) 
&=G_{nn}^{r(0)}(\omega) + \mathcal{O}(\zeta_\mathbf{k}^2), 
\label{ExpandRetGreen1}
\\
G_{n\pm1,n}^r(\zeta_\mathbf{k},\theta_\mathbf{k}=0,\omega) 
&=\frac{\zeta_\mathbf{k}}{2}G_{n\pm1,n\pm1}^{r(0)}(\omega)  G_{nn}^{r(0)}(\omega) \nonumber\\
&+ \mathcal{O}(\zeta_\mathbf{k}^2),
\label{ExpandRetGreen2}
\end{align}
where
$G_{nn}^{r(0)}(\omega) = [\hbar(\omega + n\Omega) + i\Gamma/2 - \Sigma_{nn}^r(\omega)]^{-1}$. 
Note that all other components vanish in this order.

\begin{figure*}
\centering
\includegraphics[width=0.70\textwidth]
{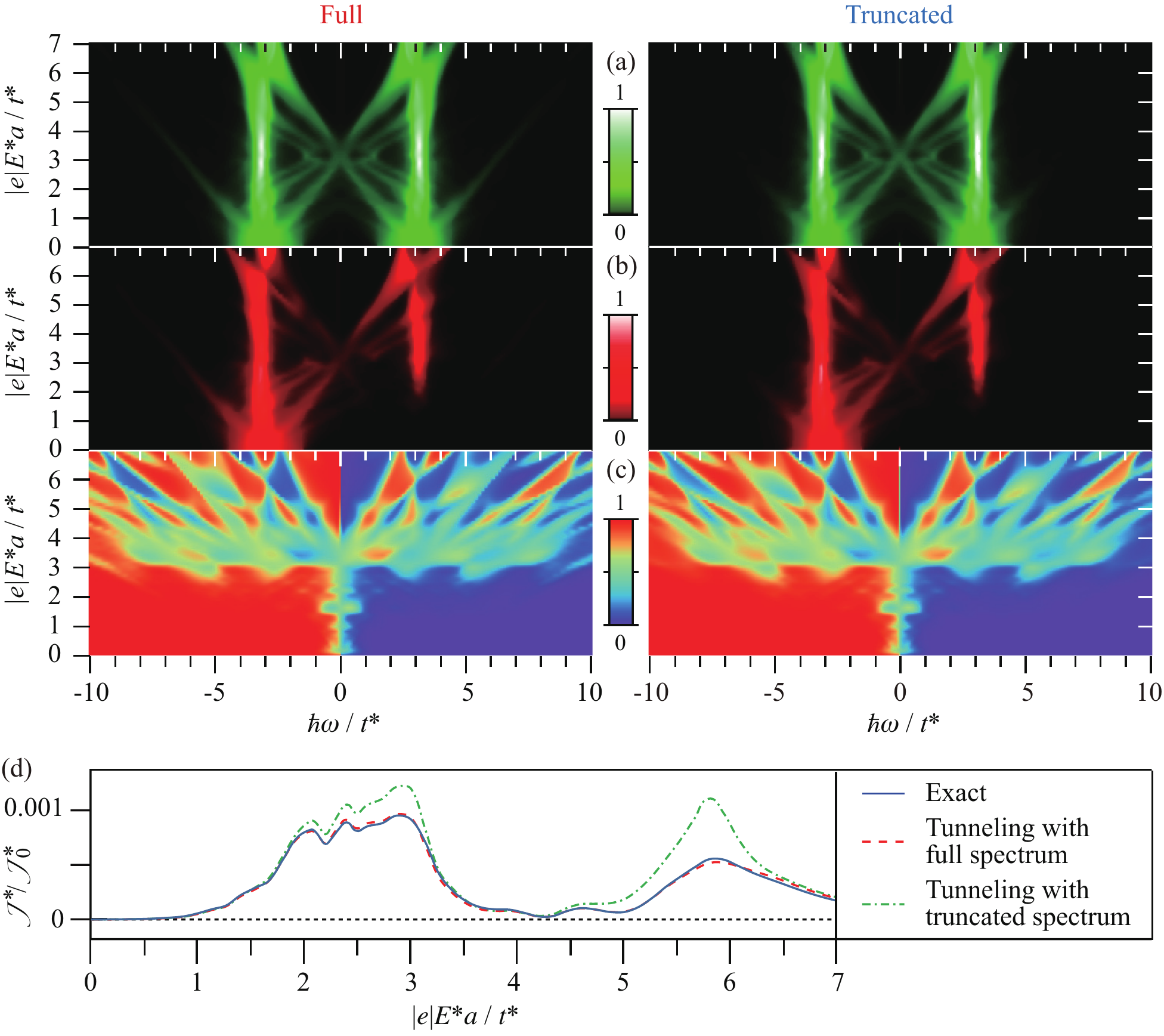} \\
\caption{
Comparison between the local (a) DOS, (b) occupation number, and (c) distribution function obtained from the full (left panels) and the zeroth-order truncated (right panels) computation.
Note that the zeroth-order truncated computation is valid in the weak-tunneling limit, i.~e. $t^* \ll \hbar\Omega$, $U$, $V_{\rm imp}$.
Panel (d) shows the comparison between the normalized direct-current density, ${\cal J}^*/{\cal J}^*_0$, obtained from the exact formula, Eq.~\eqref{ExactCurrent}, (blue solid line) and that from the tunneling formula with the full spectra, Eq.~\eqref{eq:tunneling} or \eqref{TunForm_Full}, (red dashed line) and with the zeroth-order truncated spectra, Eq.~\eqref{TunForm_Truncated}, (green dash-dotted line). 
Here, all parameters are chosen to be the same as in Fig.~4~(b); that is to say,
$U/t^*=6$, $V_{\rm imp}/t^*=0.2$, and $\Gamma/t^*=0.01$.
}
\label{Fig7_Appen}
\end{figure*}

Now, the lesser Green's function can be expanded by applying the Langreth theorem~\cite{Haug08} to the right-hand sides of Eqs.~\eqref{ExpandRetGreen1} and \eqref{ExpandRetGreen2} as follows:
\begin{align}
& G_{n+1,n}^<(\zeta_\mathbf{k},\theta_\mathbf{k}=0,\omega) \nonumber\\
& = \frac{\zeta_\mathbf{k}}{2} 
\Big\{ G_{n+1,n+1}^{r(0)}(\omega) G_{nn}^{<(0)}(\omega) + G_{n+1,n+1}^{<(0)}(\omega) G_{nn}^{a(0)}(\omega) \Big\} \nonumber\\
& ~~~ + \mathcal{O}(\zeta_\mathbf{k}^2) 
\nonumber\\
& = i\pi\zeta_\mathbf{k} \big\{ G_{n+1,n+1}^{r(0)}(\omega) \rho^{(0)}_\mathrm{loc}(\omega + n\Omega) f^{(0)}_\mathrm{loc}(\omega + n\Omega) \nonumber\\
& ~~~ + \rho^{(0)}_\mathrm{loc}(\omega + (n+1)\Omega) f^{(0)}_\mathrm{loc}(\omega + (n+1)\Omega) G_{nn}^{a(0)}(\omega) \big\} \nonumber\\
& ~~~ + \mathcal{O}(\zeta_\mathbf{k}^2),
\label{ExpandLesGreen}
\end{align}
where, according to Eq.~\eqref{KeldyshEq_main},
\begin{align}
G_{nn}^{<(0)}(\omega) 
& = G_{nn}^{r(0)}(\omega) \big[i\Gamma f_\mathrm{FD}(\omega+n\Omega) + \Sigma_{nn}^<(\omega)\big] G_{nn}^{a(0)}(\omega) 
\nonumber \\
& = -2i \mathrm{Im}G_{nn}^{r(0)}(\omega) \frac{\Gamma f_\mathrm{FD}(\omega+n\Omega) + \mathrm{Im}\Sigma_{nn}^<(\omega)}{\Gamma - 2\mathrm{Im}\Sigma_{nn}^r(\omega)} \nonumber\\
& = 2\pi i \rho_\mathrm{loc}^{(0)}(\omega + n\Omega) f_\mathrm{loc}^{(0)}(\omega + n\Omega) ,
\label{G^<(0)_nn}
\end{align}
which defines the zeroth-order local DOS and distribution function as follows:
\begin{align}
\rho^{(0)}_\mathrm{loc}(\omega+n\Omega) 
&\equiv - \frac{1}{\pi} \mathrm{Im}G_{nn}^{r(0)}(\omega),
\end{align}
and 
\begin{align}
f^{(0)}_\mathrm{loc}(\omega+n\Omega) 
&= \frac{\Gamma f_\mathrm{FD}(\omega+n\Omega) + \mathrm{Im}\Sigma^<(\omega+n\Omega)}{\Gamma - 2\mathrm{Im}\Sigma^r(\omega+n\Omega)}
\nonumber \\
&\equiv - \frac{\mathrm{Im}G_{nn}^{<(0)}(\omega)}{ 2\mathrm{Im}G_{nn}^{r(0)}(\omega)}  .
\end{align}
Note that, in the second line of Eq.~\eqref{G^<(0)_nn},  we use the fact that $\Sigma^<(\omega)$ is pure imaginary since the lesser local Green's function should be pure imaginary, which is confirmed by our numerical calculations. 
Inserting Eq.~\eqref{ExpandLesGreen} into the exact current formula in Eq.~\eqref{ExactCurrent} gives rise to the tunneling formula~\cite{Comment_SL}:
\begin{align}
\mathcal{J}_\mathrm{tun}^{*(0)} 
= &~ \frac{\mathcal{J}_0^*}{4} \int_{-\infty}^{\infty} d(\hbar\omega/t^*)~t^* \rho^{(0)}_\mathrm{loc}(\omega)~t^* \rho^{(0)}_\mathrm{loc}(\omega+\Omega) 
\nonumber \\
&\times [f^{(0)}_\mathrm{loc}(\omega) - f^{(0)}_\mathrm{loc}(\omega + \Omega)],
\label{TunForm_Truncated}
\end{align}
where ${\cal J}^*_0$ is determined to be $4\pi|e|at^*/\hbar$ by using the identity that $\int_0^\infty d\zeta ~ \zeta^3 \rho_{\mathrm{non}}(\zeta) = {t^*}^2/(2\pi)$.

Strictly speaking, the tunneling formula mentioned in the main text is different from Eq.~\eqref{TunForm_Truncated} in the sense that the full local DOS, $\rho_{\rm loc}(\omega)$, and distribution function, $f_{\rm loc}(\omega)$, are used instead of the zeroth-order truncated counterparts, $\rho^{(0)}_{\rm loc}(\omega)$ and $f^{(0)}_{\rm loc}(\omega)$; that is to say, 
\begin{align}
\mathcal{J}_\mathrm{tun}^{*} 
= &~ \frac{\mathcal{J}_0^*}{4} \int_{-\infty}^{\infty} d(\hbar\omega/t^*)~t^* \rho_\mathrm{loc}(\omega)~t^* \rho_\mathrm{loc}(\omega+\Omega) 
\nonumber \\
&\times
[f_\mathrm{loc}(\omega) - f_\mathrm{loc}(\omega + \Omega)],
\label{TunForm_Full}
\end{align}
where
\begin{align}
\rho_\mathrm{loc}(\omega+n\Omega) 
& \equiv - \frac{1}{\pi} \mathrm{Im}G_{nn}^{r}(\omega),
\\
f_\mathrm{loc}(\omega+n\Omega) 
& \equiv - \frac{\mathrm{Im}G_{nn}^{<}(\omega)}{ 2\mathrm{Im}G_{nn}^{r}(\omega)}.
\end{align}

Figure~\ref{Fig7_Appen} (a)-(c) shows the comparison between the local DOS, occupation number, and distribution function obtained from the full (left panels) and the zeroth-order truncated (right panels) computation, respectively.
As one can see, there are minor differences between the full and the truncated spectra, and therefore both spectra produce essentially the same answer.
Intriguingly, however, it is observed in Fig.~\ref{Fig7_Appen}~(d) that the tunneling formula produces much more accurate results with full spectra than with the zeroth-order truncated spectra.



\begin{thebibliography}{99}
\bibitem{Comment_G} 
In general, there are four Green's functions: the retarded, $G^r$, the advanced, $G^a$, the lesser, $G^<$, and the greater Green's function, $G^>$, accompanied by a constraint, $G^r - G^a = G^> - G^<$, which reduces the number of independent Green's functions to three.     
For a steady state, $G^r$ is related to $G^a$ via Hermitian conjugation in the frequency space, which further reduces the number to two.
\bibitem{Comment_LM} 
In the self-consistency routine, we have used the linear mixing technique to enhance the convergence of a solution. In this technique, the solution at each iteration step is linearly mixed with that at the previous one, before plugged into the next:  $G_\mathrm{mix} = (1 - \alpha) G_\mathrm{old} + \alpha G_\mathrm{new}$, where $0 < \alpha < 1$.
\bibitem{Comment_SL} 
A similar derivation can be found in Appendix B of Ref.~\cite{Wacker02} for the semiconductor superlattice.
\end{thebibliography}
\end{document}